\documentclass[aps,prc,reprint,showpacs,floatfix,superscriptaddress]{revtex4-1}

\usepackage{amsfonts,amssymb,amsmath}
\usepackage{graphicx}
\usepackage{dcolumn}
\usepackage{bm}
\usepackage[dvips]{color}
\usepackage[colorlinks=true,linkcolor=blue,citecolor=blue,urlcolor=blue]{hyperref}
\hypersetup{
  pdftitle={From nucleon-nucleon interaction matrix elements in momentum space to an operator representation},%
  pdfauthor={Dennis Weber <d.weber@gsi.de>}} 

\usepackage[english]{babel}

\begin{document}
\newcolumntype{.}{D{.}{.}{-1}}

\newcommand{\bra}[1]{\big< \,{#1}\, \big| }
\newcommand{\ket}[1]{\big| \,{#1}\, \big> }
\newcommand{\qstate}[2]{Q^{(#2)};J^\pi M #1}
\newcommand{\Q}[2]{Q^{(#1)};J^\pi #2}
\newcommand{\braket}[2]{\big< \,{#1}\, \big| \,{#2}\, \big> }

\newcommand{\expect}[1]{\big< \, {#1} \, \big>}

\newcommand{\matrixe}[3]{\big< \,{#1}\, \big| \,{#2}\, \big| \,{#3}\, \big> }


\newcommand{\op}[1]{\mathbf{#1}}

\newcommand{\ak}{a_k}
\newcommand{\ack}{\ak^{\star}}
\newcommand{\al}{a_l}
\newcommand{\acl}{\al^{\star}}
\newcommand{\am}{a_m}
\newcommand{\an}{a_n}

\newcommand{\bk}{\vec{b}_k}
\newcommand{\bck}{\bk^{\star}}
\newcommand{\bl}{\vec{b}_l}
\newcommand{\bcl}{\bl^{\star}}
\newcommand{\bn}{\vec{b}_n}

\newcommand{\Rkl}{R_{kl}}
\newcommand{\Rkm}{R_{km}}

\newcommand{\lambdakl}{\lambda_{kl}}
\newcommand{\lambdakm}{\lambda_{km}}
\newcommand{\lambdaln}{\lambda_{ln}}
\newcommand{\lambdaklmn}{\lambda_{klmn}}
\newcommand{\alphakl}{\alpha_{kl}}
\newcommand{\alphakm}{\alpha_{km}}
\newcommand{\alphaln}{\alpha_{ln}}
\newcommand{\alphaklmn}{\alpha_{klmn}}
\newcommand{\alphaklkl}{\alpha_{klkl}}
\newcommand{\pikl}{\vec{\pi}_{kl}}
\newcommand{\pikm}{\vec{\pi}_{km}}
\newcommand{\piln}{\vec{\pi}_{ln}}
\newcommand{\piklmn}{\vec{\pi}_{klmn}}
\newcommand{\rhokl}{\vec{\rho}_{kl}}
\newcommand{\rhokm}{\vec{\rho}_{km}}
\newcommand{\rholn}{\vec{\rho}_{ln}}
\newcommand{\rhoklmn}{\vec{\rho}_{klmn}}
\newcommand{\thetaklmn}{\theta_{klmn}}
\newcommand{\betaklmn}{\beta_{klmn}}

\title{From nucleon-nucleon interaction matrix elements in momentum space to an operator representation}
\author{D.~\surname{Weber}}
\affiliation{ExtreMe Matter Institute EMMI and Research Division} 
\affiliation{GSI Helmholtzzentrum f\"ur Schwerionenforschung GmbH, Planckstra{\ss}e 1, 64291
Darmstadt, Germany}
\author{H.~\surname{Feldmeier}}
\affiliation{GSI Helmholtzzentrum f\"ur Schwerionenforschung GmbH, Planckstra{\ss}e 1, 64291
Darmstadt, Germany}
\affiliation{Frankfurt Institute for Advanced Studies, Max-von-Laue-Stra{\ss}e 1, 60438 Frankfurt, Germany}
\author{H.~\surname{Hergert}}
\affiliation{Department of Physics, Ohio State University, Columbus, Ohio 43210, USA
}
\author{T.~\surname{Neff}}
\affiliation{GSI Helmholtzzentrum f\"ur Schwerionenforschung GmbH, Planckstra{\ss}e 1, 64291
Darmstadt, Germany}

\date{\today}

\begin{abstract}
Starting from the matrix elements of the nucleon-nucleon interaction in momentum space we present a method to derive an operator representation with a minimal set of operators that is required to provide an optimal description of the partial waves with low angular momentum. As a first application we use this method to obtain an operator representation for the Argonne potential transformed by means of the unitary correlation operator method and discuss the necessity of including momentum dependent operators. The resulting operator representation leads to the same results as the original momentum space matrix elements when applied to the two-nucleon system and various light nuclei. For applications in fermionic and antisymmetrized molecular dynamics, where an operator representation of a soft but realistic effective interaction is indispensable, a simplified version using a reduced set of operators is given.
\end{abstract}

\maketitle

\section{Introduction}


In recent years various realistic nucleon-nucleon (NN) potentials have been developed, such as the Argonne~V18 potential \cite{PhysRevC.51.38}, the CD~Bonn potential \cite{PhysRevC.63.024001} and the so called chiral potentials \cite{PhysRevC.68.041001,PhysRevC.66.064001,Evgeny2006654}. They all succeed in describing the experimental two-nucleon data with the same precision but differ in their off-shell behaviour, which manifests itself for example in different three-body forces needed for ab initio calculations of nuclei. Furthermore, various transformation techniques, like the renormalization group $\rm{V}_{\rm{low k}}$ \cite{Bogner20031,Bogner2003265}, the unitary correlation operator method (UCOM) \cite{Feldmeier199861,Neff2003311,Roth20043,PhysRevC.72.034002,Roth201050} and the similarity renormalization group approach (SRG) \cite{SRG2,PhysRevD.48.5863,PhysRevC.75.061001,Roth201050}, have been applied to derive effective interactions for nuclear ab initio calculations. Most of these effective realistic interactions are formulated in matrix element representation, which restricts their use to many-body treatments based on a pre-determined basis representation, such as the no-core shell model (NCSM) \cite{PhysRevC.61.044001,shellmodeltechniques,PhysRevC.64.051301,PhysRevC.79.064324,PhysRevLett.107.072501}. Thus, these interactions are not usable for many-body methods that require an explicit operator representation of the NN potential, such as fermionic molecular dynamics (FMD) \cite{FMD1,RevModPhys.72.655,Neff200869,Neffdipl} or antisymmetrized molecular dynamics (AMD) \cite{PTP.87.1185,KanadaEn'yo2003497}.

We present a method which allows to derive an approximate operator representation starting from the partial wave matrix elements of the interaction. For that purpose an ansatz for the operator representation is chosen. The unknown parameters in the ansatz are obtained from a fit to the partial wave matrix elements of the potential one aims to describe. The operators used in the ansatz will depend on the interaction under investigation. The main reason for that is the momentum-dependence of the interaction. For example, the UCOM interactions based on local potentials have a momentum dependence that is polynomial in the momenta whereas SRG interactions have a more complicated momentum dependence. In this work we consider the UCOM transformed Argonne potential. Its exact matrix elements in momentum space as well as its operator representation (in a controlled approximation) are available. The existing operator representation is very complex, but with the procedure presented in this paper, a simplified operator representation can be obtained. Whereas the full operator representation provides essentially exact matrix elements for all momenta and all partial waves, we ``only'' require our simplified operator representation to reproduce the exact matrix elements with high-precision for low momenta and in the lowest partial waves that are relevant for the description of light nuclei. In the ansatz for the operator representation we therefore use an optimized set of operators which is a subset of the operators of the exact UCOM potential. But not only the operator structure is simplified, also the radial dependencies of the interaction components are simplified in the sense that the low momentum matrix elements are not sensitive to variations on a short length scale. This is similar to the idea of coarse-grained potentials \cite{PhysRevC.88.064002}. The simplified operator representation also provides some better understanding about the importance of individual terms in the effective interaction, especially with respect to its momentum dependence.

In Sec.~\ref{sec:UCOM} the concept of UCOM and the operator representation and the partial wave matrix elements of the UCOM transformed Argonne potential are presented. Sec.~\ref{sec:op} focuses on the construction of an operator representation from the partial wave matrix elements of the interaction. We discuss the choice of a reduced set of operators to accurately describe the UCOM transformed Argonne potential for low angular momenta. In Sec.~\ref{sec:twofewsystems} we show results of calculations for two- and few-nucleon systems to demonstrate that the operator representation with the reduced set of operators indeed leads to the same results as the exact, but more complex, UCOM transformed Argonne potential. 

\section{The unitary correlation operator method (UCOM)\label{sec:UCOM}}
\subsection{Concept}
The concept of the Unitary Correlation Operator Method \cite{Roth201050} is to imprint the short-range central and tensor correlations induced by the nuclear interaction \cite{PhysRevC.84.054003} on ``simple'' many-body states $\ket{\Psi}$, such as Slater determinants. The unitary operator $\op{C}$ describes the transformation between the many-body state $\ket{\Psi}$, which possibly contains long-range but no short-range central and tensor correlations, and the state $\ket{\widehat{\Psi}}$ that contains all correlations:

\begin{equation}
\ket{\widehat{\Psi}} = \op{C} \, \ket{\Psi}.
\label{eq:unitaryoperator}
\end{equation}
To calculate the matrix element $\bra{\widehat{\Psi}}\op{B}\ket{\widehat{\Psi}'}$ of an operator $\op{B}$ one can either work with the bare operator $\op{B}$ and correlated states $\ket{\widehat{\Psi}}$ or use a correlated operator

\begin{equation}
\op{\widehat{B}} = \op{C}^{-1} \op{B} \, \op{C} =\op{C}^{\dag} \op{B} \, \op{C} 
\label{eq:correlatedoperator}
\end{equation}
and uncorrelated states $\ket{\Psi}$ instead:

\begin{equation}
\bra{\widehat{\Psi}}\op{B}\ket{\widehat{\Psi}'} = \bra{\Psi} \op{C}^{\dag} \op{B} \, \op{C} \ket{\Psi'} = \bra{\Psi}\op{\widehat{B}}\ket{\Psi'}.
\label{eq:correlatedoperator2}
\end{equation}
Both methods are equivalent, but it is generally more convenient to work with uncorrelated simple states and the correlated operators.

The correlation operator $\op{C}$ is decomposed into the unitary operators $\op{C}_{\Omega}$ and $\op{C}_r$ describing the tensor and radial correlations, respectively:
\begin{eqnarray}
\op{C} = \op{C}_{\Omega} \op{C}_r.
\label{eq:correlationseperated}
\end{eqnarray}
The following ansatz with hermitian two-body generators $\op{g}_r$ and $\op{g}_{\Omega}$ is used:
\begin{eqnarray}
\op{C}_{\Omega} = \mbox{exp}\Big\{-i\sum_{i<j}{\op{g}_{\Omega,ij}}\Big\}, \hspace{0.3cm} \op{C}_{r} = \mbox{exp}\Big\{-i\sum_{i<j}{\op{g}_{r,ij}}\Big\}. 
\label{eq:correlationgenerators}
\end{eqnarray}
The form of these generators reflects the structure of the central and tensor correlations.

\subsection {UCOM generators}
The short-range repulsion of the NN interaction prevents the nucleons from approaching each other closer than the extent of the repulsive core. That means the two-body density at short relative distances will be strongly suppressed in the correlated many-body state. This effect can be achieved by a distance-dependent shift of the radial wave function. Using the projection of the relative momentum $\vec{\op{p}}=\frac{1}{2}(\vec{\op{p}}_1-\vec{\op{p}}_2)$ on the relative distance vector $\vec{\op{r}}=\vec{\op{r}}_1-\vec{\op{r}}_2$ in the two-body subsystem,
\begin{eqnarray}
\op{p}_r = \frac{1}{2}\Big[\vec{\op{p}}\frac{\vec{\op{r}}}{\op{r}}+\frac{\vec{\op{r}}}{\op{r}}\vec{\op{p}}\Big],
\label{eq:radmomentum}
\end{eqnarray}
and the shift function $s_{ST}(\op{r})$ which describes the amplitude of the radial shift for each spin-isospin channel, the generator can be written as
\begin{eqnarray}
\op{g}_r = \sum_{ST} \frac{1}{2} \Big[\op{p}_r s_{ST}(\op{r}) + s_{ST}(\op{r}) \op{p}_r\Big]\,\op{\Pi}_{ST},
\label{eq:centralgenerator}
\end{eqnarray}
where $\op{\Pi}_{ST}$ is a projector on spin $S$ and isospin $T$.

The tensor force induces correlations between the orientation of the total spin and that of the relative distance vector $\vec{\op{r}}$ of a pair of nucleons. These correlations can be imprinted by a tangential shift perpendicular to $\vec{\op{r}}$, generated by the ``orbital momentum'' operator
\begin{eqnarray}
\vec{\op{p}}_{\Omega} = \vec{\op{p}}-\frac{\vec{\op{r}}}{\op{r}}\op{p}_r.
\label{eq:orbitalmomentum}
\end{eqnarray}
The generator of the tensor correlation operator is given by
\begin{eqnarray}
\op{g}_{\Omega} = \sum_{T}  \vartheta_{T}(\op{r}) S_{12}(\vec{\op{r}},\vec{\op{p}_{\Omega}})\,\op{\Pi}_{1T},
\label{eq:tensorgenerator}
\end{eqnarray}
where the function $\vartheta_{T}(\op{r})$ describes the distance-dependence of the angle of the tangential shift in order to orient the relative distance vector more parallel ($T$=$0$) or anti-parallel ($T$=$1$) to the total spin direction ($S$=$1$) of the particle pair.  The generating operator $S_{12}(\vec{\op{r}},\vec{\op{q}_{\Omega}})$ is given by the general definition 
\begin{eqnarray}S_{12}(\vec{\op{a}},\vec{\op{b}})&=&\frac{3}{2}\Big[(\vec{\op{\sigma}}_1\cdot\vec{\op{a}})(\vec{\op{\sigma}}_2\cdot\vec{\op{b}})+(\vec{\op{\sigma}}_1\cdot\vec{\op{b}})(\vec{\op{\sigma}}_2\cdot\vec{\op{a}})\Big] \nonumber \\ &&-\frac{1}{2}(\vec{\op{\sigma}}_1\cdot\vec{\op{\sigma}}_2)(\vec{\op{a}}\cdot\vec{\op{b}}+\vec{\op{b}}\cdot\vec{\op{a}}).
\end{eqnarray}

The correlation functions $s_{ST}(\op{r})$ and $\vartheta_T(\op{r})$ are specific to the used NN interaction. They can be derived for example by performing an energy minimization in the lowest angular momentum channels (UCOM(var)) \cite{Neff2003311,Roth20043}. An alternative method is to extract them from a SRG transformation (UCOM(SRG)) \cite{PhysRevC.77.064003}. In this article, we use the UCOM(SRG) transformed potentials. As the operator structure for both types of UCOM interactions is the same, the presented method to derive the operator representation from the matrix elements of the interaction can be used for UCOM(var) potentials as well.

\subsection {The UCOM potential \label{sec:UCOMop}}
The formalism described above is applied to correlate the nuclear Hamiltonian.

Because the correlation operators are A-body operators, the UCOM transformation of any operator $\op{B}$ contains irreducible contributions from up to A-particle operators as well:
\begin{eqnarray}
\op{C}^{\dag}\op{B}\op{C} = \op{\widehat{B}}=\op{\widehat{B}}^{[1]}+\op{\widehat{B}}^{[2]}+\op{\widehat{B}}^{[3]}+\cdots+\op{\widehat{B}}^{[A]},
\end{eqnarray}
where  $\op{\widehat{B}}^{[n]}$ stands for the irreducible $n$-body part \cite{Feldmeier199861}. The importance of the contributions with higher particle numbers $n$ is however expected to be small since the range of the UCOM transformation is by construction small compared to the mean interparticle distance in nuclei. Since the treatment of terms with higher $n$ in many-body calculations is difficult, we restrict ourselves in the following to the two-body approximation (denoted by C2), where all contributions above the two-body level are discarded:
\begin{eqnarray}
\op{\widehat{B}}^{C2}=\op{\widehat{B}}^{[1]}+\op{\widehat{B}}^{[2]}.
\end{eqnarray}
To obtain the UCOM potential we start from an uncorrelated two-body Hamiltonian
\begin{eqnarray}
\op{H}= \op{T} + \op{V},
\label{eq:hamiltonian}
\end{eqnarray}
where the kinetic energy $\op{T}=\op{T}_{\rm{c.m.}}+\op{T}_{\mathrm{int}}$ contains the center of mass term  $\op{T}_{\rm{c.m.}}$ and the intrinsic part $\op{T}_{\mathrm{int}}$. $\op{V}$ is a realistic two-body potential. In this paper we use the Argonne~V18 potential \cite{PhysRevC.51.38} without charge dependent terms. It can be written as a sum of radial functions $v^{P}_{ST}(\op{r})$ which depend only on the relative distance, multiplied with the corresponding operators 
\begin{eqnarray} \op{\mathcal{O}}_P\in\{\op{1},\;\vec{\op{L}}^{\,2},\;(\vec{\op{L}}\cdot\vec{\op{S}}),\;\op{S}_{12},\;S_{12}(\vec{\op{L}},\vec{\op{L}})\}
\label{eq:operators}
\end{eqnarray}
that act on angular momentum and spin degrees of freedom:
\begin{eqnarray}
\op{V}_{\rm{Argonne}} &=& \sum_P \sum_{ST} v^P_{ST}(\op{r})\op{\mathcal{O}}_P\, \op{\Pi}_{ST}. 
\label{eq:argonnepotential0}
\end{eqnarray}
$\op{S}_{12}$ stands for $S_{12}(\frac{\vec{\op{r}}}{\op{r}},\frac{\vec{\op{r}}}{\op{r}})$ and we replace the quadratic spin-orbit operator $(\vec{\op{L}}\cdot\vec{\op{S}})^2$ used in Ref.~\cite{PhysRevC.51.38} by the tensor operator $S_{12}(\vec{\op{L}},\vec{\op{L}})$ by means of the relation 
\begin{eqnarray}
 (\vec{\op{L}}\cdot\vec{\op{S}})^2=\frac{1}{6}S_{12}(\vec{\op{L}},\vec{\op{L}})+\frac{2}{3}\vec{\op{L}}^{\,2}\op{\Pi}_{S=1}-\frac{1}{2}(\vec{\op{L}}\cdot\vec{\op{S}}).\label{eq:ls2}
\end{eqnarray}
The Argonne potential then reads explicitly
\begin{eqnarray}
\op{V}_{\rm{Argonne}} &=& \sum_{ST} v^C_{ST}(\op{r}) \,\op{\Pi}_{ST} \nonumber \\
&+&\sum_{ST} v^{L2}_{ST}(\op{r})\vec{\op{L}}^{\,2}\,\op{\Pi}_{ST} \nonumber \\
&+& \sum_{T} v^{LS}_{1T}(\op{r})(\vec{\op{L}}\cdot\vec{\op{S}})\,\op{\Pi}_{1T} \nonumber \\ &+& \sum_{T}v^T_{1T}(\op{r})\op{S}_{12} \,\op{\Pi}_{1T} \nonumber \\ 
&+& \sum_{T} v^{Tll}_{1T}(\op{r})S_{12}(\vec{\op{L}},\vec{\op{L}}) \,\op{\Pi}_{1T}.
\label{eq:argonnepotential}
\end{eqnarray}

\subsubsection{Correlated Hamiltonian}
In two-body approximation the UCOM potential $\op{V}_{\rm{UCOM}}$ is defined as the two-body part of the correlated Hamiltonian
\begin{eqnarray} 
\op{C}_r^{\dag}\op{C}_{\Omega}^{\dag}\op{H}\op{C}_{\Omega}\op{C}_r  &\stackrel{C2}{=}& \op{\widehat{T}}^{[1]}+ \op{\widehat{T}}^{[2]}+ \op{\widehat{V}}^{[2]}\nonumber\\
&=:& \op{\widehat{T}}^{[1]}+\op{V}_{\rm{UCOM}}.
\end{eqnarray}

To obtain the radially correlated Hamiltonian $\op{C}_r^{\dag}\op{H}\op{C}_r$, we have to calculate the correlated kinetic energy $\op{C}_r^{\dag}\op{T}\op{C}_r$ and the correlated potential $\op{C}_r^{\dag}\op{V}\op{C}_r$. Since the generator $\op{g}_r$ commutes with the operators $\op{\mathcal{O}}_P$ occurring in the Argonne potential, only the correlated radial functions have to be calculated \cite{Feldmeier199861}:
\begin{eqnarray} 
\op{C}_r^{\dag}\left(v(\op{r})\op{\mathcal{O}}_P\right)\op{C}_r &=& v(\op{C}_r^{\dag}\op{r}\op{C}_r)\,\op{\mathcal{O}}_P \nonumber \\
&=&v(R_+(\op{r}))\,\op{\mathcal{O}}_P=:\widehat{v}(\op{r})\op{\mathcal{O}}_P.
\end{eqnarray}
The correlation function $R_+(\op{r})$ is specific for each $ST$-channel and connected to the shift function $s(\op{r})$ in Eq.~(\ref{eq:centralgenerator}) by the relation $\int_r^{R_+(r)}\frac{\mbox{d}\xi}{s(\xi)}=1$.
The radial correlation of the kinetic energy creates one- and two-body terms:
\begin{eqnarray} 
\op{C}_r^{\dag}\op{T}\op{C}_r =\op{\widehat{T}}^{[1]}+\op{\widehat{T}}^{[2]}=\op{T}+\op{\widehat{T}}^{[2]}.
\end{eqnarray}
While the one-body part $\op{\widehat{T}}^{[1]}$ is identical to the uncorrelated kinetic energy operator $\op{T}$, the two-body term $\op{\widehat{T}}^{[2]}$ contains, besides a central and an $\vec{\op{L}}^{\,2}$-operator term, a quadratic momentum contribution:
\begin{eqnarray} 
\op{\widehat{T}}^{[2]}&=&W(\op{r})+\left(\frac{1}{2\mu_{\Omega}(\op{r})}-\frac{1}{2\mu_{r}(\op{r})}\right)\frac{\vec{\op{L}}^{\,2}}{\op{r}^2}\nonumber\\
&+&\frac{1}{2}\left(\vec{\op{p}}^{\,2}\frac{1}{2\mu_{r}(\op{r})} + \frac{1}{2\mu_{r}(\op{r})}\vec{\op{p}}^{\,2}\right),
\end{eqnarray}
where $W(\op{r})$, $\mu_{r}(\op{r})$ and $\mu_{\Omega}(\op{r})$ are functions of the correlation function $R_+(\op{r})$ and its derivatives:
\begin{subequations} 
\begin{eqnarray} 
W(\op{r})&=&\frac{7R_+''(\op{r})^2}{4R_+'(\op{r})^4}-\frac{R_+'''(\op{r})}{2R_+'(\op{r})^3}\\
\frac{1}{2\mu_{r}(\op{r})}&=&\frac{1}{2\mu}\left(R_+'(\op{r})^{-2}-1\right)\\
\frac{1}{2\mu_{\Omega}(\op{r})}&=&\frac{1}{2\mu}\left(\op{r}^2R_+(\op{r})^{-2}-1\right).
\end{eqnarray}
\end{subequations}
The transformation of the Hamiltonian with the tensor correlation operator $\op{C}_{\Omega}$, Eq.~(\ref{eq:correlationgenerators}), can be evaluated by means of the Baker-Campbell-Hausdorff expansion:
\begin{eqnarray} 
\op{C}_{\Omega}^{\dag}\op{H}\op{C}_{\Omega} =\op{H} + i\big[\op{g}_{\Omega},\op{H}\big]+\frac{i^2}{2}\big[\op{g}_{\Omega},\big[\op{g}_{\Omega},\op{H}\big]\big]+\cdots.
\label{eq:commutator}
\end{eqnarray}
For the radial part of the intrinsic kinetic energy, $\op{T}_r=\frac{\op{p}_r^2}{2\mu}$, the expansion Eq.~(\ref{eq:commutator}) terminates after the first order and creates, besides additional spin-orbit and tensor contributions, a term containing the momentum dependent tensor operator $S_{12}(\vec{\op{r}},\vec{\op{p}_{\Omega}})$ \cite{Neff2003311,Roth20043}:
\begin{eqnarray} 
\op{C}_{\Omega}^{\dag}\op{T}_r\op{C}_{\Omega}&=&\op{T}_r-\frac{1}{2\mu}\Big[\left(\op{p}_r\vartheta'(\op{r})+\vartheta'(\op{r})\op{p}_r\right)S_{12}(\vec{\op{r}},\vec{\op{p}_{\Omega}})\nonumber \\
&&+9\left(\vartheta'(\op{r})\right)^2 \left(\vec{\op{S}}^2+3(\vec{\op{L}}\cdot\vec{\op{S}})+(\vec{\op{L}}\cdot\vec{\op{S}})^2\right)\Big],\nonumber \\
\end{eqnarray}
where $(\vec{\op{L}}\cdot\vec{\op{S}})^2$ can be expanded via Eq.~(\ref{eq:ls2}). 

The commutator of $\op{g}_{\Omega}$ and the other operators in Eq.~(\ref{eq:argonnepotential}), $\vec{\op{L}}^{\,2},\,(\vec{\op{L}}\cdot\vec{\op{S}}),\,\op{S}_{12}$ and $S_{12}(\vec{\op{L}},\vec{\op{L}})$, creates additional central, spin-orbit and tensor contributions as well as terms with the new operator $\bar{S}_{12}(\vec{\op{p}}_{\Omega},\vec{\op{p}}_{\Omega})=2\op{r}^2S_{12}(\vec{\op{p}}_{\Omega},\vec{\op{p}}_{\Omega})+S_{12}(\vec{\op{L}},\vec{\op{L}})-\frac{1}{2}\op{S}_{12}$. The commutator algebra is not closed and higher orders of the expansion Eq.~(\ref{eq:commutator}) lead to additional new operators with the structure $\vec{\op{L}}^{\,2n}(\vec{\op{L}}\cdot\vec{\op{S}})$,  $\vec{\op{L}}^{\,2n}\op{S}_{12}(\vec{\op{L}},\vec{\op{L}})$ and $\vec{\op{L}}^{\,2n}\bar{S}_{12}(\vec{\op{p}}_{\Omega},\vec{\op{p}}_{\Omega})$, with $n=1,\,2,\,3,\,\cdots$. Due to the centrifugal barrier, the relative wave function is more and more suppressed at short relative distances as $L$ increases, and its overlap with $\vartheta_{T}(\op{r})$ becomes progressively smaller as well. Therefore, $\op{C}_{\Omega}$ essentially reduces to the identity operator and one may perform a partial summation in Eq.~(\ref{eq:commutator}) neglecting all terms beyond the third order in angular momentum $\vec{\op{L}}$ \cite{Roth20043}.

\subsubsection{Operator representation}
For the initial Argonne V18 interaction, the UCOM potential has the structure \cite{PhysRevC.72.034002}
\begin{eqnarray} 
\op{V}_{\rm{UCOM}} &=&\sum_{ST} V^C_{ST}(\op{r})\,\op{\Pi}_{ST} \nonumber \\
&+&\sum_{ST}  V^{L2}_{ST}(\op{r})\vec{\op{L}}^{\,2}\,\op{\Pi}_{ST} \nonumber \\
&+& \sum_{ST} \frac{1}{2}\big[\vec{\op{p}}^{\,2}V_{ST}^{p2}(\op{r}) + V_{ST}^{p2}(\op{r})\vec{\op{p}}^{\,2}\big]\,\op{\Pi}_{ST} \nonumber \\
&+& \sum_{T} \,V^{LS}_{1T}(\op{r})(\vec{\op{L}}\cdot\vec{\op{S}})\,\op{\Pi}_{1T} \nonumber \\ 
&+& \sum_{T} \,V^{L2LS}_{1T}(\op{r})\vec{\op{L}}^{\,2}(\vec{\op{L}}\cdot\vec{\op{S}})\,\op{\Pi}_{1T} \nonumber \\ 
&+& \sum_{T}V^T_{1T}(\op{r})\op{S}_{12} \,\op{\Pi}_{1T} \nonumber \\
&+& \sum_{T} V^{Tll}_{1T}(\op{r})S_{12}(\vec{\op{L}},\vec{\op{L}}) \,\op{\Pi}_{1T} \nonumber \\
&+& \sum_{T}V^{Tpp}_{1T}(\op{r})\bar{S}_{12}(\vec{\op{p}}_{\Omega},\vec{\op{p}}_{\Omega}) \,\op{\Pi}_{1T} \nonumber \\
&+& \sum_{T} \frac{1}{2}V_{1T}^{L2Tpp}(\op{r})\cdot \nonumber \\
&&\big[\vec{\op{L}}^{\,2}\bar{S}_{12}(\vec{\op{p}}_{\Omega},\vec{\op{p}}_{\Omega}) + \bar{S}_{12}(\vec{\op{p}}_{\Omega},\vec{\op{p}}_{\Omega}) \vec{\op{L}}^{\,2}\big]\,\op{\Pi}_{1T} \nonumber\\ 
&+& \sum_{T} \frac{1}{2}\big[\op{p}_rV_{1T}^{Trp}(\op{r}) + V_{1T}^{Trp}(\op{r})\op{p}_r\big]\cdot\nonumber \\
&&S_{12}(\vec{\op{r}},\vec{\op{p}_{\Omega}})\,\op{\Pi}_{1T}, 
\label{eq:ucompotential} 
\end{eqnarray}
with new radial functions $V^{P}_{ST}(\op{r})$ that depend on the initial potential and the correlation functions $s_{ST}(\op{r})$ and $\vartheta_T(\op{r})$. 

Compared to the bare potential, given in Eq.~(\ref{eq:argonnepotential}), the operator structure is more complicated and contains the additional spin-orbit and tensor operators $\vec{\op{L}}^{\,2}(\vec{\op{L}}\cdot\vec{\op{S}})$, $\bar{S}_{12}(\vec{\op{p}}_{\Omega},\vec{\op{p}}_{\Omega})$ and the explicitly momentum dependent operators $\vec{\op{p}}^{\,2}$ and $\op{p}_rS_{12}(\vec{\op{r}},\vec{\op{p}_{\Omega}})$. This momentum dependence originates from the UCOM transformation of the kinetic energy operator and leads to a different operator structure as is obtained in a momentum space representation of the bare Argonne potential \cite{PhysRevC.84.034003}. Therefore, the UCOM potential $\op{V}_{\rm{UCOM}}$ is always nonlocal, even if the initial interaction is local.

\subsubsection {Matrix element representation \label{sec:UCOMme}}

In matrix representation with basis states that possess good angular momentum and spin quantum numbers, like in the shell model, the operator structure and its truncation is not explicitly needed. In that case it is possible to perform the UCOM transformation exactly in the given basis, e.g. the partial wave basis in momentum space $\ket{k(LS)J;T}$ with relative momentum $k$, relative angular momentum $L$, total spin $S$, total angular momentum $J$ and total isospin $T$. The expressions for the matrix elements of the UCOM transformed kinetic energy and the Argonne potential are given in App.~\ref{sec:appendixUCOMmes},  Eqs.~(\ref{eq:ucommat}).

By using these relations the matrix elements of $\op{V}_{\rm{UCOM}}$ can be calculated:
\begin{multline}
\bra{k(LS)J;T} \op{V}_{\rm{UCOM}}\ket{k'(L'S)J;T} = \\
 \bra{k(LS)J;T} \op{C}_r^{\dag}\op{C}_{\Omega}^{\dag}\op{H}\op{C}_{\Omega}\op{C}_r-\op{T}\ket{k'(L'S)J;T}.\label{eq:vucomme} 
\end{multline}
These matrix elements are exact on the two-body level and include no approximations like the partial summation of the Baker-Campbell-Hausdorff expansion in the operator representation Eq.~\eqref{eq:ucompotential}.

\subsection{Correlation functions}
In this work we use the UCOM(SRG) transformation \cite{PhysRevC.77.064003,Roth201050} obtained with the SRG flow parameter of $\alpha=0.04\,\rm{fm}^4$ (UCOM(0.04)). Whereas the interaction for this flow parameter is much softer than the original Argonne interaction and can be used, for example, in NCSM calculations, it is still not soft enough for many-body approaches like FMD and AMD. We therefore also discuss the UCOM(SRG) transformed Argonne potential with $\alpha=0.2\,\rm{fm}^4$ (UCOM(0.20)). For this larger flow parameter the tensor force in the UCOM(SRG) interaction is significantly weaker and therefore better adapted to the FMD and AMD model spaces. For both selected values of the flow parameter the UCOM(SRG) interaction in two-body approximation gives binding energies for $^3\rm{H}$ and $^4\rm{He}$ that are close to the experimental values \cite{Roth201050}. 
It should be emphasized that different flow parameters in the UCOM(SRG) transformation yield the same operator structure Eq.~(\ref{eq:ucompotential}), but different radial functions. Consequently, the method discussed in the following section is applicable for UCOM(SRG) transformations with any flow parameter $\alpha$.

\section{Operator representation determination starting from UCOM matrix elements \label{sec:op}}
The UCOM transformed Argonne potential is one of the few effective realistic potentials for which the operator representation as well as the matrix element representation is known. As shown in Sec.~\ref{sec:UCOMop}, the operator representation of the UCOM transformed Argonne potential, given by Eq.~(\ref{eq:ucompotential}), is more complicated than the one of the initial Argonne potential Eq.~(\ref{eq:argonnepotential}), even though terms with higher powers of the angular momentum operator are neglected. In a previous work it has been shown that these neglected terms are not important \cite{Roth20043}. The question arises if one can reduce the number of operators further without loosing accuracy.

In the following we present a method to derive an operator representation for an interaction that is given by its partial wave matrix elements. This method is applied to obtain a simpler operator representation for the UCOM transformed Argonne potential that nevertheless reproduces the matrix elements of the partial wave channels with low angular momenta with the same accuracy as the exact UCOM transformed Argonne matrix elements.

\subsection{Method \label{sec:method}}
We start from the partial wave matrix elements of the NN potential, here the UCOM transformed Argonne potential $\op{V}_{\mathrm{UCOM}}$. We define an ansatz for the operator representation to describe these matrix elements:
\begin{eqnarray} 
\op{V}_{\rm{ansatz}}=\sum_P\sum_{ST}\frac{1}{2}\left[\op{\mathcal{O}}_P \mathcal{V}^P_{ST}(\op{r})+\mathcal{V}^P_{ST}(\op{r})\op{\mathcal{O}}_P\right]\,\op{\Pi}_{ST},\nonumber \\
\label{eq:ansatz}
\end{eqnarray}  
with operators $\op{\mathcal{O}}_P$ and corresponding radial functions $\mathcal{V}^P_{ST}(\op{r})$, which are parameterized, for convenient use in FMD as described in Sec.~\ref{sec:FMD}, by a sum of Gaussians:
\begin{eqnarray} 
\mathcal{V}^P_{ST}(r)=\sum_{\mu}r^{n_P}\gamma^{P}_{ST,\mu}\mbox{exp}\left\{-\frac{r^2}{2\kappa_{\mu}}\right\},
\label{eq:parameterization}
\end{eqnarray}
where $n_P=0,\,2,\,3$ depending on the operator $\op{\mathcal{O}}_P$ (see Eqs.~(\ref{eq:FMDrad})). 

Next, we choose a set of parameters $\kappa$ that covers the range of the NN interaction. In this work a geometrical sequence 
\begin{eqnarray*}
\kappa_{\mu}=\kappa_1 \cdot b^{\mu-1}
\end{eqnarray*}
with $\kappa_1=0.05\,\rm{fm}^2$ and $b=2$ is used, so that 
\begin{eqnarray*} 
\kappa=\left\{0.05,\, 0.1,\,0.2,\,\cdots,\,6.4\right\}\mbox{fm}^2.
\end{eqnarray*}
Width parameters larger than $6.4\,\rm{fm}^2$ are not required since the interaction has a range of only a few $\mathrm{fm}$. It is only necessary to include width parameters starting from $\kappa_1=0.05\,\rm{fm}^2$ (which corresponds to momentum transfers up to roughly $1/\sqrt{\kappa_1}\approx 4.5\,\rm{fm}^{-1}$), because short relative distances (corresponding to large momentum transfers) are not resolved at energies relevant for nuclear structure.
  
The parameters $\gamma^{P}_{ST,\mu}$ are obtained by a fit of the ansatz to the partial wave matrix elements of $\op{V}_{\rm{UCOM}}$. For that purpose we have to derive the partial wave matrix elements of the ansatz for the operator representation:
\begin{align} 
&\bra{k(LS)J;T}\op{V}_{\rm{ansatz}}\ket{k'(L'S)J;T}=\nonumber \\
&\sum_P\! \bra{\!k(LS)J;T\!}\frac{1}{2}\!\left[\!\op{\mathcal{O}}_P \mathcal{V}^P_{ST}(\op{r})\!+\!\mathcal{V}^P_{ST}(\op{r})\op{\mathcal{O}}_P\!\right]\!\ket{\!k'(L'S)J;T\!}.
\label{eq:ansatzme}
\end{align}  
In case of operators $\op{\mathcal{O}}_P$ acting only in angular momentum and spin space (e.g., all the operators occurring in the Argonne potential Eq.~(\ref{eq:operators}) and $\vec{\op{L}}^{\,2}(\vec{\op{L}}\cdot\vec{\op{S}})$ and $\bar{S}_{12}(\vec{\op{p}}_{\Omega},\vec{\op{p}}_{\Omega})$ in the UCOM potential Eq.~(\ref{eq:ucompotential})) $\op{\mathcal{O}}_P$ and its radial function $\mathcal{V}^P_{ST}(\op{r})$ commute. Using the parameterization Eq.~(\ref{eq:parameterization}), we find
\begin{subequations}
\begin{align} 
\bra{k(LS)J;T}\mathcal{V}^P_{ST}(\op{r})\op{\mathcal{O}}_P&\ket{k'(L'S)J;T}=\nonumber \\
=\frac{2}{\pi}\int_0^{\infty}dr r^2 j_L(kr)&\mathcal{V}^P_{ST}(r)j_{L'}(k'r)\cdot\nonumber \\ 
&\bra{(LS)J;T} \op{\mathcal{O}}_P \ket{(L'S)J;T}\nonumber \\
=\sum_{\mu}\gamma^{P}_{ST,\mu}\frac{2}{\pi}\int_0^{\infty}drr^2 &j_L(kr)\,r^{n_P}\mbox{exp}\left\{-\frac{r^2}{2\kappa_{\mu}}\right\}j_{L'}(k'r)\cdot\nonumber \\ 
&\bra{(LS)J;T} \op{\mathcal{O}}_P \ket{(L'S)J;T}.\label{eq:ansatzme2}
\end{align}  
One has to calculate the matrix elements of the operator $\op{\mathcal{O}}_P$ and the integrals over the parameterized radial functions, which have analytical solutions for the Gaussian parameterization (see App.~\ref{sec:radint}). The required matrix elements $\bra{(LS)J;T} \op{\mathcal{O}}_P \ket{(L'S)J;T}$ are given in App.~\ref{sec:opme}. For the momentum dependent terms containing the operators $\vec{\op{p}}^{\,2}$ and $\op{p}_rS_{12}(\vec{\op{r}},\vec{\op{p}_{\Omega}})$, one finds
\begin{align} 
\bra{k(LS)J;T}\frac{1}{2}\left[\vec{\op{p}}^{\,2}\mathcal{V}_{ST}^{p2}(\op{r}) + \mathcal{V}_{ST}^{p2}(\op{r})\vec{\op{p}}^{\,2}\right]\ket{k'(L'S)J;T}=\nonumber \\
\sum_{\mu}\gamma^{p2}_{ST,\mu}\frac{2}{\pi}\int_0^{\infty}dr r^2 j_L(kr)\mbox{exp}\left\{-\frac{r^2}{2\kappa_{\mu}}\right\}j_{L'}(k'r)\cdot\nonumber \\ 
\frac{1}{2}(k^2+k'^2)\delta_{LL'} \label{eq:ansatzmemom}
\end{align}  
and
\begin{align} 
\bra{\!k(LS)J;T\!}\frac{1}{2}\!\left[\op{p}_r\mathcal{V}_{1T}^{Trp}(\op{r})\! + \! \mathrm{h. c.}\right]\!S_{12}(\vec{\op{r}},\vec{\op{p}_{\Omega}})\!\ket{\!k'(L'S)J;T\!}=\nonumber \\
\sum_{\mu}\!\gamma^{Trp}_{ST,\mu}\frac{i}{\pi}\Bigg[\!\int_0^{\infty}dr r^2 \!\!\left(\!\frac{\partial}{\partial r}rj_L(kr)\!\right)\!r^2\mbox{exp}\!\left\{-\frac{r^2}{2\kappa_{\mu}}\right\}\!j_{L'}(k'r)\nonumber \\
-\!\int_0^{\infty}dr r^2 j_L(kr)\,r^2\mbox{exp}\!\left\{-\frac{r^2}{2\kappa_{\mu}}\right\}\left(\!\frac{\partial}{\partial r}rj_{L'}(k'r)\!\right)\!\!\Bigg]\cdot\nonumber 
\\ \bra{k(LS)J;T}S_{12}(\vec{\op{r}},\vec{\op{p}_{\Omega}})\ket{k'(L'S)J;T}.
\label{eq:ansatzmemom2}
\end{align}
\label{eq:ansatzmemoms}
\end{subequations}
Like for Eq.~(\ref{eq:ansatzme2}), the radial integrals and the matrix elements of the operators can be calculated analytically (see App.~\ref{sec:radint}), so that one obtains also analytical expressions for the matrix elements containing momentum operators.

A fit of this expression to the exact matrix elements of $\op{V}_{\rm{UCOM}}$ determines the optimal parameters $\gamma^{P}_{ST,\mu}$ and thereby the radial functions Eq.~(\ref{eq:parameterization}).

For $S=0$ the potential Eq.~(\ref{eq:ucompotential}) contains only contributions from operators that have no dependence on the spin operator, namely $\op{1}$, $\vec{\op{L}}^2$ and $\vec{\op{p}}^{\,2}$. In this case we fit the ansatz directly to the partial wave matrix elements $\bra{k(L0)L;T}\op{V}\ket{k'(L0)L;T}$. For $S=1$ the potential contains besides the central part also spin-orbit and tensor contributions. The spin-orbit part contains all operators of tensor rank one in spin space (like $\vec{\op{L}}\cdot\vec{\op{S}}$ and $\vec{\op{L}}^2(\vec{\op{L}}\cdot\vec{\op{S}})$) and the tensor part contains the operators of tensor rank two (for example $\op{S}_{12}$ and $S_{12}(\vec{\op{L}},\vec{\op{L}})$). By using appropriate linear combinations of the matrix elements of the potential with given angular momentum $L$ and different total angular momenta $J$, one can separate the central, spin-orbit and tensor contributions. Details of this method can be found in App.~\ref{sec:separate}. We use these linear combinations of the matrix elements to fit the central, spin-orbit, and tensor part independently.

\subsection{Choice of the operators}
The choice of the operators $\op{\mathcal{O}}_P$ in Eq.~(\ref{eq:ansatz}) plays a crucial role for the quality of the fitted operator representation. The success of a certain set of operators in describing the interaction matrix elements depends on the potential under consideration. In this work, we focus on the UCOM transformed Argonne potential and a set of operators providing a good description of the matrix elements of this potential.

\begin{figure*}[ht!]
\includegraphics[width=0.85\textwidth]{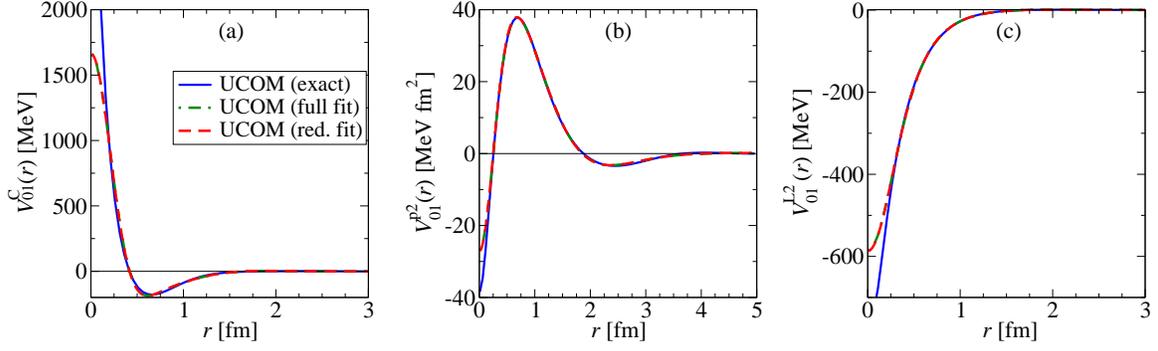}
\caption{\label{fig:ucomfit}(Color online) Radial functions of the exact UCOM(0.04) potential (Eq.~(\ref{eq:ucompotential})) (blue solid line) compared to those of the fit with the full set of operators (green dashed-dotted) and the fit with the reduced set of operators (red dashed). The latter two curves coincide at the used resolution. {(a): $S=0$, $T=1$ central}; (b): $S=0$, $T=1$ central momentum dependent; (c): $S=0$, $T=1$ central quadratic angular momentum dependent.}
\end{figure*}

In order to test if the fitting procedure contains enough momentum space matrix elements to fix uniquely the parameters $\gamma^{P}_{ST,\mu}$, a first ansatz that includes all operators which occur in the operator representation of the UCOM potential (denoted by ``full UCOM'') is used. This corresponds to the operator set no.~(1) in Tab.~\ref{tab:choice}. Taking the exact matrix elements Eqs.~(\ref{eq:ucommat}) of the UCOM transformed Argonne potential with momenta $k$ up to $10\,\mathrm{fm}^{-1}$ (which is large enough to guarantee no undesired effects outside the fitting region for the used minimal Gaussian width parameter $\kappa_1=0.05\,\rm{fm}^2$) and angular momenta $L$ up to $4$ as input, the radial functions $\mathcal{V}_{ST}^{P}(r)$ obtained by the fit agree with the exact radial functions $V_{ST}^{P}(r)$ of the exact UCOM potential $\op{V}_{\rm{UCOM}}$ \cite{Weber}.
As an example, Fig.~\ref{fig:ucomfit} compares for $S=1$, $T=0$ the central radial functions $V_{01}^C(r)$ with $\mathcal{V}_{01}^C(r)$, the central momentum dependent $V_{01}^{p2}(r)$ with $\mathcal{V}_{01}^{p2}(r)$ and the quadratic angular momentum dependent $V_{01}^{L2}(r)$ with $\mathcal{V}_{01}^{L2}(r)$. Small deviations occur only at short relative distances, which due to the volume element $r^2\mathrm{d}r$ do not contribute to the low-momentum matrix elements.

The fact that the fit to the exact matrix elements yields the same radial functions as those calculated analytically with the UCOM transformation shows that the method to obtain operators from a set of matrix elements is working. It also shows that the exact matrix elements Eqs.~(\ref{eq:ucommat}) and the already truncated set of operators that neglect higher powers in $\vec{\op{L}}^2$ originating from the Baker-Campbell-Hausdorff expansion are equivalent.

As a next step, we consider a reduced ansatz for the operator representation with only a subset of the operators contained in Eq.~(\ref{eq:ucompotential}). Again, the radial functions are obtained from the fit to the exact partial wave matrix elements of the UCOM transformed Argonne potential. Obviously, a fit with less operators will not reproduce all matrix elements of the potential with the same quality. For a given set of angular momenta $L=0,\,1,\cdots,L_{\mathrm{max}}$ the set of operators $\op{\mathcal{O}}_P$ provides a number of linearly independent matrices in each $ST$-channel. By increasing this number, one can of course increase the quality of fitting the momentum space partial wave matrix elements for $L=0,\,1,\cdots,L_{\mathrm{max}}$  because of the increased number of radial functions. However, just as in a polynomial fit, the obtained interactions might get worse outside the fitted domain for $L>L_{\mathrm{max}}$. As discussed before, it helps that for high angular momenta $L$ the centrifugal barrier becomes dominant and the contributions from the interaction are getting smaller in comparison. Thus, we optimize the fit for angular momenta relevant in nuclear structure and afterwards make sure (by testing the operator representation in many-body calculations, see Sec.~\ref{sec:twofewsystems}) that the deviations in the matrix elements with higher $L$ do not affect the calculated two-nucleon phase shifts and the properties of other light nuclear systems in a significant fashion.

Because there is no unique choice of a reduced set of operators describing correctly the lowest angular momentum channels, it is advisable to use as few operators as necessary to maintain correct matrix elements and nuclear properties. 

\begin{figure*}[ht]
\centering
\vspace{4ex}
\includegraphics[width=0.59\textwidth]{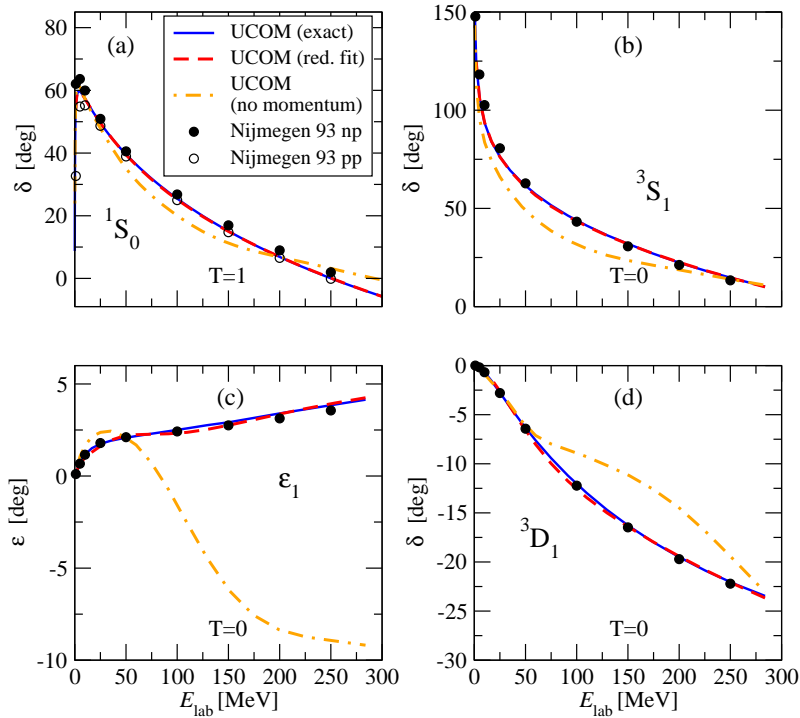}
\caption{\label{fig:mom}(Color online) Nucleon-nucleon phase shifts and mixing angle calculated with different potentials. Blue solid line: exact UCOM(0.04) matrix elements. Red dashed line: reduced UCOM(0.04) fit with set no.~(3) in Tab.~\ref{tab:choice}. Orange dash-dotted line: UCOM(0.04) fit excluding explicitly momentum dependent operators (corresponding to set no.~(2) in Tab.~\ref{tab:choice}). The dots indicate the results of the 1993 Nijmegen partial wave analysis \cite{PhysRevC.48.792}.}
\end{figure*}

An overview of possible sets of operators is given in Tab.~\ref{tab:choice}. It turns out that it is necessary to include the momentum dependent operators  $\vec{\op{p}}^{\,2}$ and $\op{p}_rS_{12}(\vec{\op{r}},\vec{\op{p}_{\Omega}})$, which originate from the correlated kinetic energy, present in the set of operators. The contributions from the momentum dependent terms can not be absorbed by modifying the radial functions of the other operators because they are local and therefore not able to describe this momentum dependence. The quadratic momentum dependence is a characteristic feature of the UCOM potential. These terms replace the strong short range repulsion and the short range tensor which are responsible for the undesired scattering to high momentum states. Therefore, one cannot omit them in the reduced set of operators. Fig.~\ref{fig:mom} shows the phase shifts calculated with a refitted operator representation excluding the momentum dependent operators (corresponding to set no.~(2) in Tab.~\ref{tab:choice}). From these results it is obvious that it is not possible to maintain the correct phase shifts of the exact UCOM matrix elements without using momentum dependent operators in the ansatz.

\begin{table}[h]
\centering
\begin{ruledtabular}
\begin{tabular}{c|cccccccccc||c}
set & \multicolumn{10}{c||}{operator structure}&pw's\\
no. &$C$&$L2$&$p2$&$LS$&$L2LS$&$T$&$TLL$&$Tpp$&$Trp$&$L2Tpp$& $L\leq 2$ \\ \hline
(1)&$\checkmark$&$\checkmark$&$\checkmark$&$\checkmark$&$\checkmark$&$\checkmark$&$\checkmark$&$\checkmark$&$\checkmark$&$\checkmark$&$\checkmark$ \\ \hline
(2)&$\checkmark$&$\checkmark$&$\times$&$\checkmark$&$\checkmark$&$\checkmark$&$\checkmark$&$\checkmark$&$\times$&$\checkmark$&$\times$ \\ \hline
(3)&$\checkmark$&$\checkmark$&$\checkmark$&$\checkmark$&$\times$&$\checkmark$&$\checkmark$&$\times$&$\checkmark$&$\times$&$\checkmark$ \\ \hline 
(4)&$\checkmark$&$\checkmark$&$\checkmark$&$\checkmark$&$\times$&$\checkmark$&$\times$&$\checkmark$&$\checkmark$&$\times$&$\checkmark$ \\ \hline
(5)&$\checkmark$&$\checkmark$&$\checkmark$&$\checkmark$&$\times$&$\checkmark$&$\times$&$\times$&$\checkmark$&$\times$&$\times$ 
\end{tabular}
\end{ruledtabular}
\caption{Different sets of operators used in ${V}_{\rm{ansatz}}$ (for the abbreviations see Eq.~\eqref{eq:ucompotential}). The matrix elements, phase shifts and deuteron properties calculated with the fitted interaction are compared with those obtained with the exact UCOM matrix elements. A check mark in the column ``pw's $L\leq 2$'' indicates that the set is able to reproduce the results of the exact matrix elements in the partial waves with angular momentum up to $L=2$. \label{tab:choice} }
\end{table}

\begin{figure*}[th!]
\begin{minipage}{0.31\textwidth}
\centering
{\large(a) \bf UCOM (exact)}
\end{minipage}
\begin{minipage}{0.31\textwidth}
\centering
{\large(b) \bf UCOM (reduced fit)}
\end{minipage}
\begin{minipage}{0.31\textwidth}
\centering
{\large(c) \bf difference}
\end{minipage}
\begin{minipage}{0.31\textwidth}
\centering
\includegraphics[width=\textwidth]{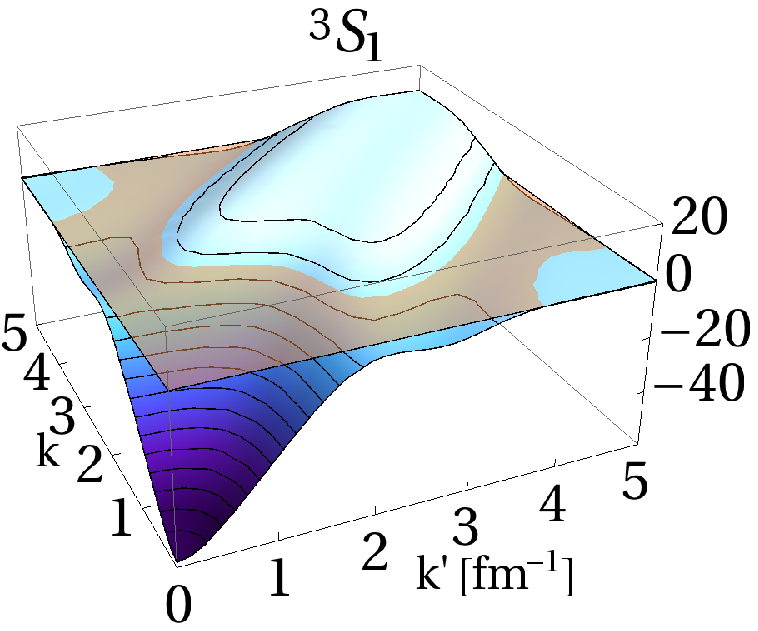}
\end{minipage}
\hspace{0.2cm}
\begin{minipage}{0.31\textwidth}
\centering
\includegraphics[width=\textwidth]{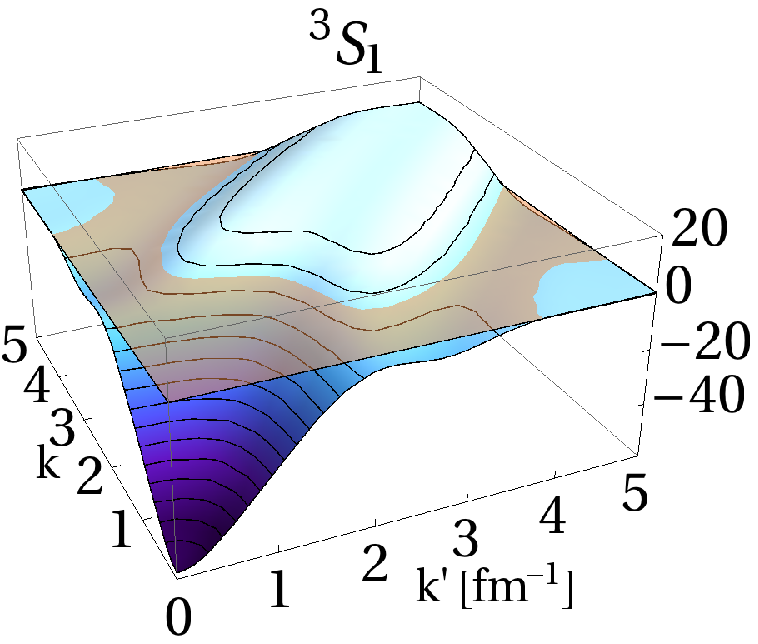}
\end{minipage}
\hspace{0.2cm}
\begin{minipage}{0.30\textwidth}
\centering
\includegraphics[width=\textwidth]{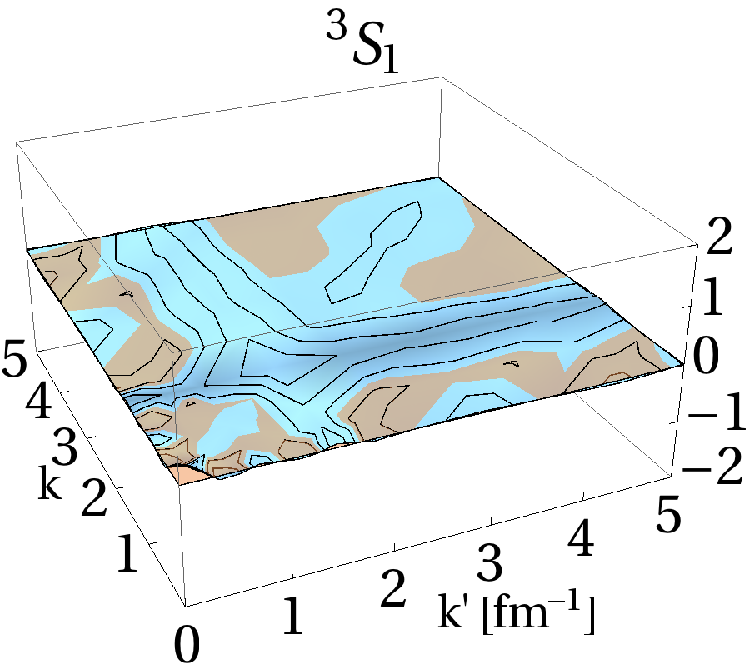}
\end{minipage}
\begin{minipage}{0.31\textwidth}
\centering
\includegraphics[width=\textwidth]{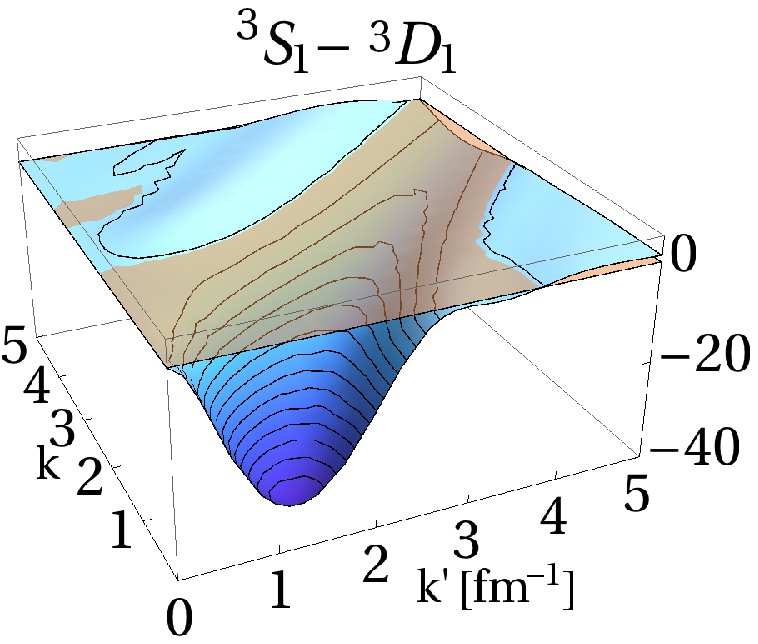}
\end{minipage}
\hspace{0.2cm}
\begin{minipage}{0.31\textwidth}
\centering
\includegraphics[width=\textwidth]{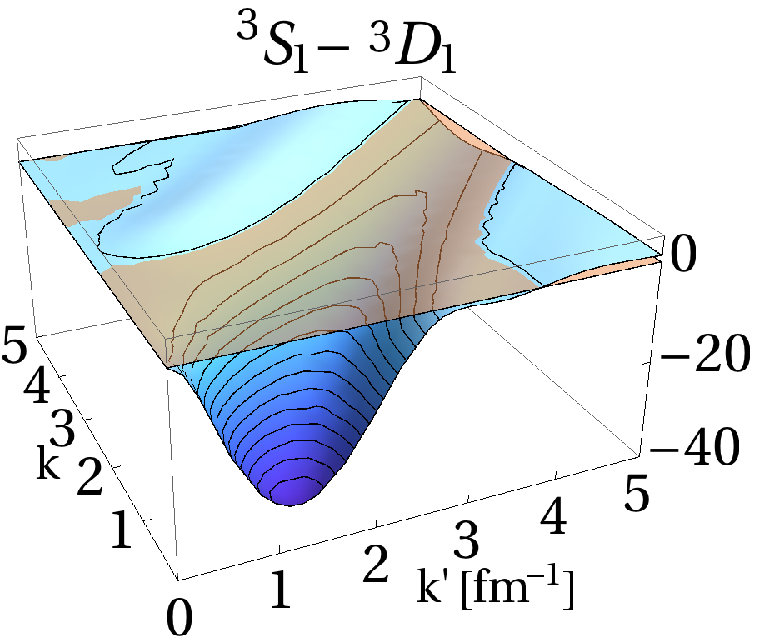}
\end{minipage}
\hspace{0.2cm}
\begin{minipage}{0.31\textwidth}
\centering
\includegraphics[width=\textwidth]{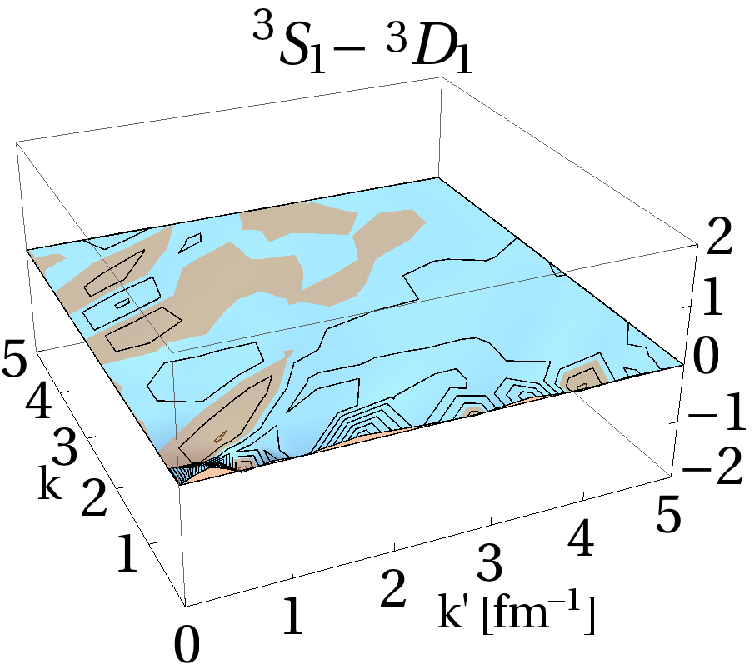}
\end{minipage}
\begin{minipage}{0.31\textwidth}
\centering
\includegraphics[width=\textwidth]{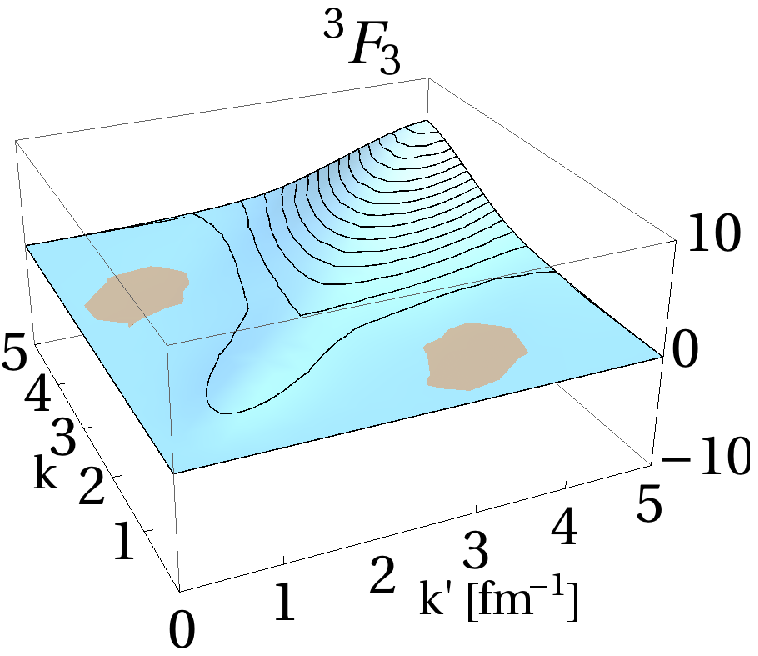}
\end{minipage}
\hspace{0.2cm}
\begin{minipage}{0.31\textwidth}
\centering
\includegraphics[width=\textwidth]{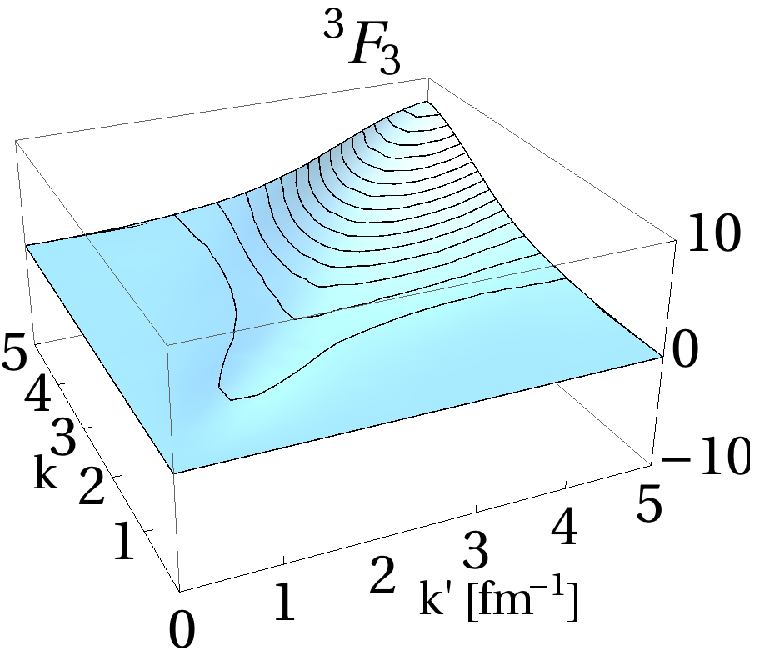}
\end{minipage}
\hspace{0.2cm}
\begin{minipage}{0.31\textwidth}
\centering
\includegraphics[width=\textwidth]{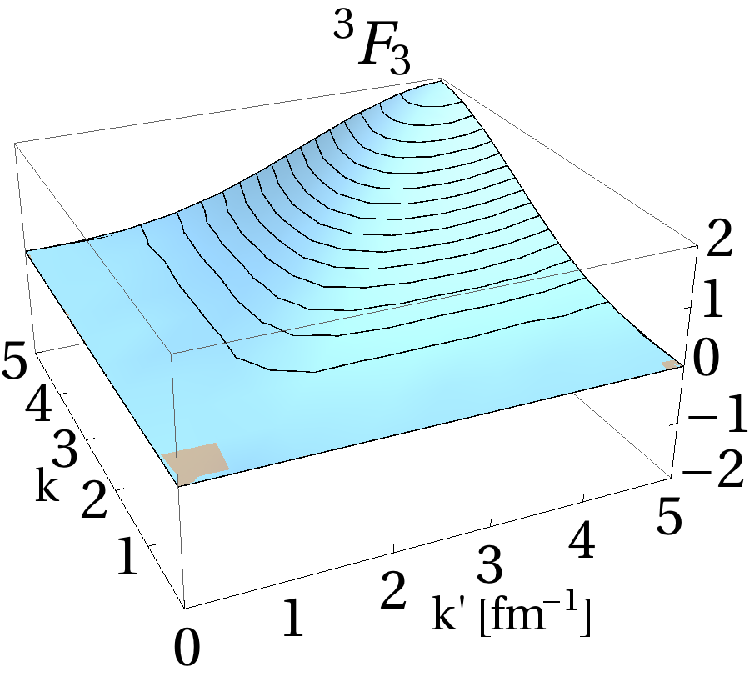}
\end{minipage}
\caption{\label{fig:me}(Color online) Exact matrix elements $\bra{k(LS)J;T)}V_{\rm{UCOM}}\ket{k'(L'S)J;T}$ (in units of $\rm{MeV}\,\rm{fm}^3$) of (a) the UCOM(0.04) potential, (b) matrix elements $\bra{k(LS)J;T)}V^{(\rm{red.})}_{\rm{UCOM}}\ket{k'(L'S)J;T}$ obtained by a fit using the reduced ansatz Eq.~(\ref{eq:reducompotential}) and (c) difference between them for the $^3\rm{S}_1$, $^3\rm{S}_1$-$^3\rm{D}_1$ and $^3\rm{F}_3$ channel. The brown plane marks the position of the zero-plane. Note the different scale for matrix elements and differences for the $^3\rm{F}_3$ wave.  }
\end{figure*}

The reduced sets no.~(3) and (4) in Tab.~\ref{tab:choice} reproduce the deuteron properties and phase shifts with angular momenta up to $L=2$. For smaller sets of operators, for example set no.~(5), the matrix elements and two nucleon properties of the fitted interaction do not agree well with those of the exact UCOM interaction even for the lowest angular momenta.

In this article, we focus on an ansatz with the reduced set of operators no.~(3), keeping in mind that other choices for the set of operators are also possible. This choice  provides a minimal set of operators which reproduces matrix elements and properties of the UCOM potential for angular momenta up to $L=2$. Ansatz no.~(3), which we call in the following ``reduced UCOM'' potential, reads explicitly:
\begin{eqnarray} 
\op{V}^{(\rm{red.})}_{\rm{UCOM}} &=&\sum_{ST} \mathcal{V}^C_{ST}(\op{r}) \,\op{\Pi}_{ST} \nonumber \\
&+&\sum_{ST} \mathcal{V}^{L2}_{ST}(\op{r})\vec{\op{L}}^{\,2}\,\op{\Pi}_{ST} \nonumber \\
&+& \sum_{ST} \frac{1}{2}\big[\vec{\op{p}}^{\,2}\mathcal{V}_{ST}^{p2}(\op{r}) + \mathcal{V}_{ST}^{p2}(\op{r})\vec{\op{p}}^{\,2}\big]\,\op{\Pi}_{ST} \nonumber \\
&+& \sum_{T} \,\mathcal{V}^{LS}_{1T}(\op{r})(\vec{\op{L}}\cdot\vec{\op{S}})\,\op{\Pi}_{1T} \nonumber \\ 
&+& \sum_{T}\mathcal{V}^T_{1T}(\op{r})\op{S}_{12} \,\op{\Pi}_{1T} \nonumber \\
&+& \sum_{T} \mathcal{V}^{Tll}_{1T}(\op{r})S_{12}(\vec{\op{L}},\vec{\op{L}}) \,\op{\Pi}_{1T} \nonumber \\
&+& \sum_{T} \frac{1}{2}\big[\op{p}_r\mathcal{V}_{1T}^{Trp}(\op{r}) + \mathcal{V}_{1T}^{Trp}(\op{r})\op{p}_r\big]\cdot\nonumber \\
&&S_{12}(\vec{\op{r}},\vec{\op{p}_{\Omega}})\op{\Pi}_{1T} . 
\label{eq:reducompotential} 
\end{eqnarray}
Compared to the UCOM potential with the full set of operators, Eq.~(\ref{eq:ucompotential}) the reduced UCOM potential Eq.~(\ref{eq:reducompotential}) lacks the operators $\vec{\op{L}}^{\,2}(\vec{\op{L}}\cdot\vec{\op{S}}),\, \bar{S}_{12}(\vec{\op{p}}_{\Omega},\vec{\op{p}}_{\Omega}),\, \vec{\op{L}}^{\,2}\bar{S}_{12}(\vec{\op{p}}_{\Omega},\vec{\op{p}}_{\Omega})$. Compared to the operator set of the initial Argonne potential Eq.~(\ref{eq:argonnepotential}) it is supplemented by the momentum dependent operators $\vec{\op{p}}^{\,2}$ and $\op{p}_rS_{12}(\vec{\op{r}},\vec{\op{p}_{\Omega}})$.

It should be noted that, besides the tensor operator $\op{S}_{12}$, the operator $S_{12}(\vec{\op{r}},\vec{\op{p}}_\Omega)$ is the only one that connects $L$ with $L\pm 2$ states \cite{Neff2003311}. Thus it is the only operator in the effective interaction that can soften the strong short range tensor correlations induced by the original tensor interaction. 

The partial wave matrix elements of this ansatz (calculated by means of Eqs.~\eqref{eq:ansatzmemoms}) are fitted to the exact partial wave matrix elements of the UCOM transformed Argonne potential. Partial wave matrix elements with momenta $k$ up to $10\,\rm{fm}^{-1}$ and angular momentum up to $L=4$ are used to perform the fit. In the fitting procedure we put different weights to the considered partial wave matrix elements (or linear combinations containing only central, spin-orbit or tensor components) in order to optimize the fit for low angular momentum channels, which we want to reproduce with high accuracy. These weights are listed in Tabs.~\ref{tab:weightreducom1} and \ref{tab:weightreducom2} in App.~\ref{Sec:Weight}.

\begin{figure*}[ht]
\centering

\includegraphics[width=0.6\textwidth]{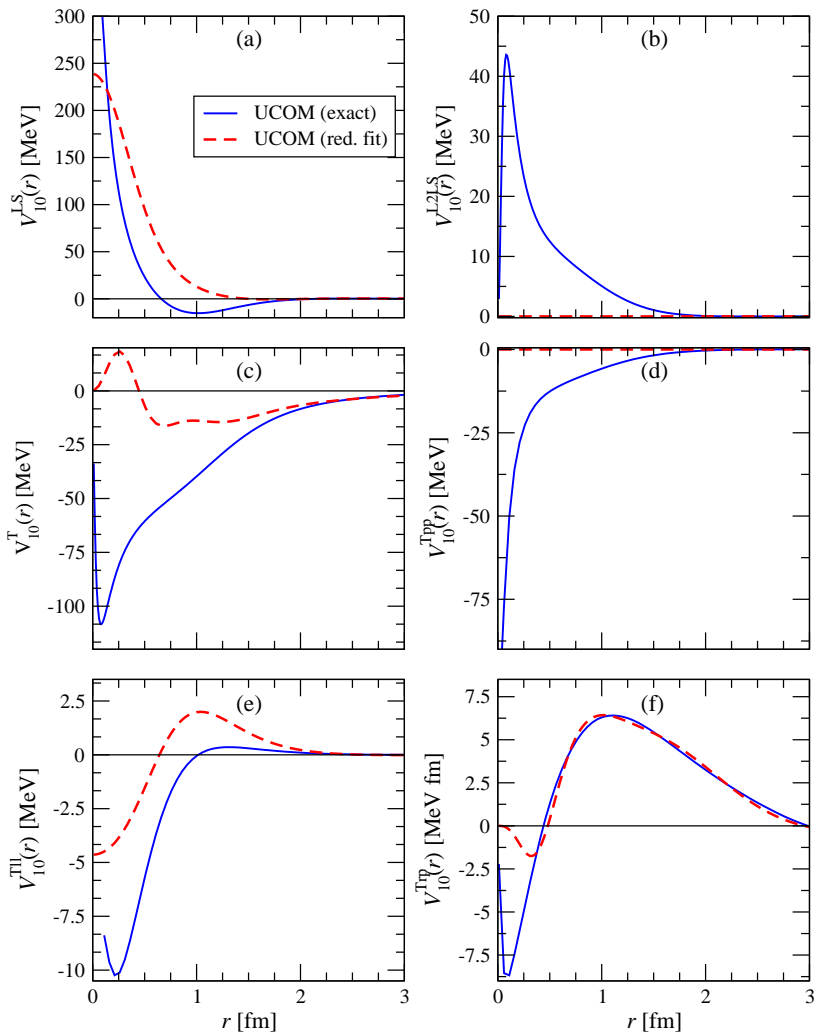}
\caption{\label{fig:ucomrad}(Color online) Radial functions of the exact UCOM(0.04) potential (blue solid line) and the fit to UCOM(0.04) matrix elements with the reduced set of operators (red dashed line) for $S=1$ and $T=0$. (a) and (b): the two spin-orbit terms; (c) to (f): the radial functions of the tensor terms.}
\end{figure*}
\begin{figure}[ht!]
\centering
\includegraphics[width=0.3\textwidth]{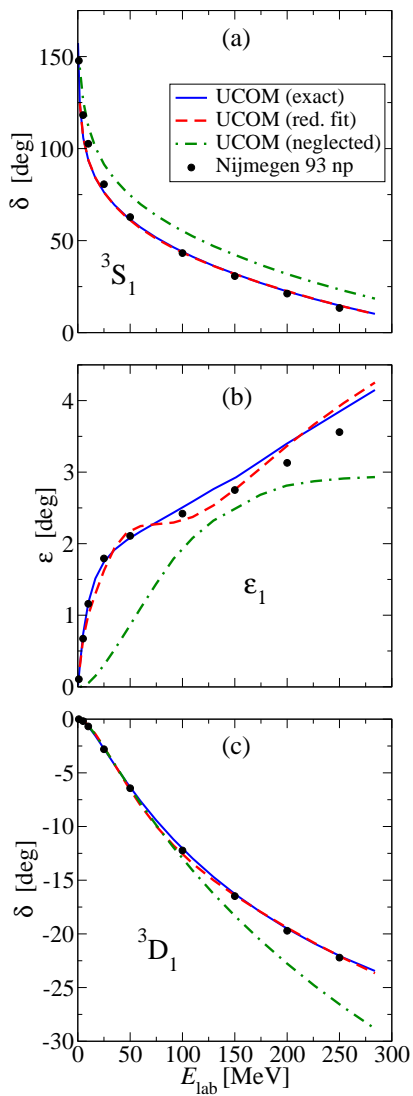}
\caption{\label{fig:neglected}(Color online) Nucleon-nucleon phase shifts calculated with the exact matrix elements of the UCOM(0.04) transformed Argonne potential (blue solid line), the reduced UCOM(0.04) fit (red dashed line) and the UCOM(0.04) transformed Argonne potential neglecting the terms with $\vec{L}^{\,2}(\vec{L}\cdot\vec{S})$, $\bar{S}_{12}(\vec{p}_{\Omega},\vec{p}_{\Omega})$ and $\vec{L}^{\,2}\bar{S}_{12}(\vec{p}_{\Omega},\vec{p}_{\Omega})$ without refitting the radial functions (green dot-dashed line). The dots indicate the results of the 1993 Nijmegen partial wave analysis \cite{PhysRevC.48.792}.
}
\end{figure}

The parameters $\gamma^P_{ST,\mu}$ of the reduced UCOM(0.04) potential, obtained from the fit to the exact UCOM(0.04) matrix elements, are listed in App.~\ref{sec:parameters}, Tabs.~\ref{tab:st00} - \ref{tab:st11}. In Tabs.~\ref{tab:2000st00} - \ref{tab:2000st11} we list the parameters for the reduced UCOM(0.20) fit. 

Fig.~\ref{fig:me} shows for selected partial waves the exact matrix elements of the UCOM(0.04) potential and the matrix elements of the reduced UCOM(0.04) fit. Because the partial waves with low angular momentum have greater weights in the fit than those with higher angular momentum, the deviations between the exact UCOM(0.04) matrix elements and the matrix elements of the reduced potential are very small in the $^3\rm{S}_1$ and $^3\rm{S}_1$-$^3\rm{D}_1$ channel. The small deviations in the $^3\mathrm{S}_1$ and $^3\mathrm{S}_1$-$^3\mathrm{D}_1$ channel are also reflected in the phase shifts displayed in Fig.~\ref{fig:neglected}. Exact UCOM(0.04) and reduced UCOM(0.04) result in the same phase shifts up to $E_{\mathrm{lab}}=300\,\rm{MeV}$. They also reproduce the experimental phase shifts (Nijmegen 1993 np \cite{PhysRevC.48.792}), which is to be expected, as UCOM is phase-shift equivalent to the initial Argonne potential by construction. The $^3\rm{F}_3$ channel exhibits deviations between the exact and the reduced UCOM(0.04) matrix elements (Fig.~\ref{fig:me}), which also shows up in deviations between the phase shifts (shown in Fig.~\ref{fig:phs2}) at higher energies.

Figs.~\ref{fig:ucomfit} and \ref{fig:ucomrad} show some of the radial functions of the reduced UCOM(0.04) potential $\op{V}^{(\rm{red.})}_{\rm{UCOM}}$ in comparison with the functions of the exact UCOM(0.04) potential $\op{V}_{\rm{UCOM}}$. Since the reduced set of operators contains the same central operator terms ($\op{1}$, $\vec{\op{L}}^{\,2}$ and $\vec{\op{p}}^{\,2}$) as the exact UCOM potential, the radial functions of the central part are the same, except for the differences at short relative distances. The radial functions $\mathcal{V}^P_{ST}(\op{r})$ of $\op{V}^{(\rm{red.})}_{\rm{UCOM}}$ of the spin-orbit and tensor terms differ from the radial functions $V^P_{ST}(\op{r})$ of $\op{V}_{\rm{UCOM}}$ since they have to compensate the missing operator terms. For instance, for $L=1$ and $2$ the spin-orbit term $\mathcal{V}^{LS}_{ST}(\op{r})$ in $\op{V}^{(\rm{red.})}_{\rm{UCOM}}$ can absorb  the contribution that comes from the neglected $\vec{\op{L}}^{\,2}(\vec{\op{L}}\cdot\vec{\op{S}})$ term such, that for 
$L$=$1$, $T$=$1$ and $L$=$2$, $T$=$0$ 
\begin{align}
&\bra{\!k(L1)J;T\!}\mathcal{V}^{LS}_{1T}(\op{r})(\vec{\op{L}}\!\cdot\!\vec{\op{S}})\ket{\!k'(L1)J;T\!}=\nonumber \\ 
&\bra{\!k(L1)J;T\!}{V}^{LS}_{1T}(\op{r})(\vec{\op{L}}\!\cdot\!\vec{\op{S}})\!+\!V^{L2LS}_{1T}(\op{r})\vec{\op{L}}^{\,2}(\vec{\op{L}}\!\cdot\!\vec{\op{S}})\ket{\!k'(L1)J;T\!} \nonumber
\end{align}
and differences occur only for $L$=$3$ and higher. The radial function $\mathcal{V}^{T}_{1T}(\op{r})$ of the tensor term in the reduced UCOM potential compensates for low $L$ the missing $\bar{S}_{12}(\vec{\op{p}}_{\Omega},\vec{\op{p}}_{\Omega})$ terms that connect $L$ with $L\pm2$ states. This modification is counterbalanced in channels with $L=L'$ by the new radial function $\mathcal{V}^{Tll}_{1T}(\op{r})$ of the tensor operator $S_{12}(\vec{\op{L}},\vec{\op{L}})$, which is diagonal in $L$. The radial function $\mathcal{V}^{Trp}_{1T}(\op{r})$ of the momentum dependent tensor operator agrees, disregarding deviations at short relative distances, with the one of the exact UCOM potential.   

If one simply neglects the omitted operators $\vec{\op{L}}^{\,2}(\vec{\op{L}}\cdot\vec{\op{S}})$, $\bar{S}_{12}(\vec{\op{p}}_{\Omega},\vec{\op{p}}_{\Omega})$ and $\vec{\op{L}}^{\,2}\bar{S}_{12}(\vec{\op{p}}_{\Omega},\vec{\op{p}}_{\Omega})$ without refitting the radial functions, one sees a substantial difference in the matrix elements and phase shifts, which is illustrated in Fig.~\ref{fig:neglected}. This means that the contribution of these operators is not small but is absorbed by the reduced set of operators due to refitting.

\section{Testing the operator representation in two- and few-nucleon systems \label{sec:twofewsystems}}

\subsection{Two-nucleon properties}
The initial Argonne potential is a realistic NN interaction: It reproduces the NN scattering phase shifts and the properties of the deuteron. By construction, the short-ranged unitary UCOM transformation does not affect these properties and thus the UCOM transformed interaction can also be regarded as an effective realistic potential. We demand that the reduced UCOM fit reproduces the deuteron properties and phase shifts with the same quality as the exact UCOM matrix elements, at least for angular momenta up to $L=2$. We use the exact UCOM matrix elements and those of the reduced UCOM fit to calculate the two-nucleon properties and compare the results.

\begin{figure*}[ht!]
\centering
\includegraphics[width=0.6\textwidth]{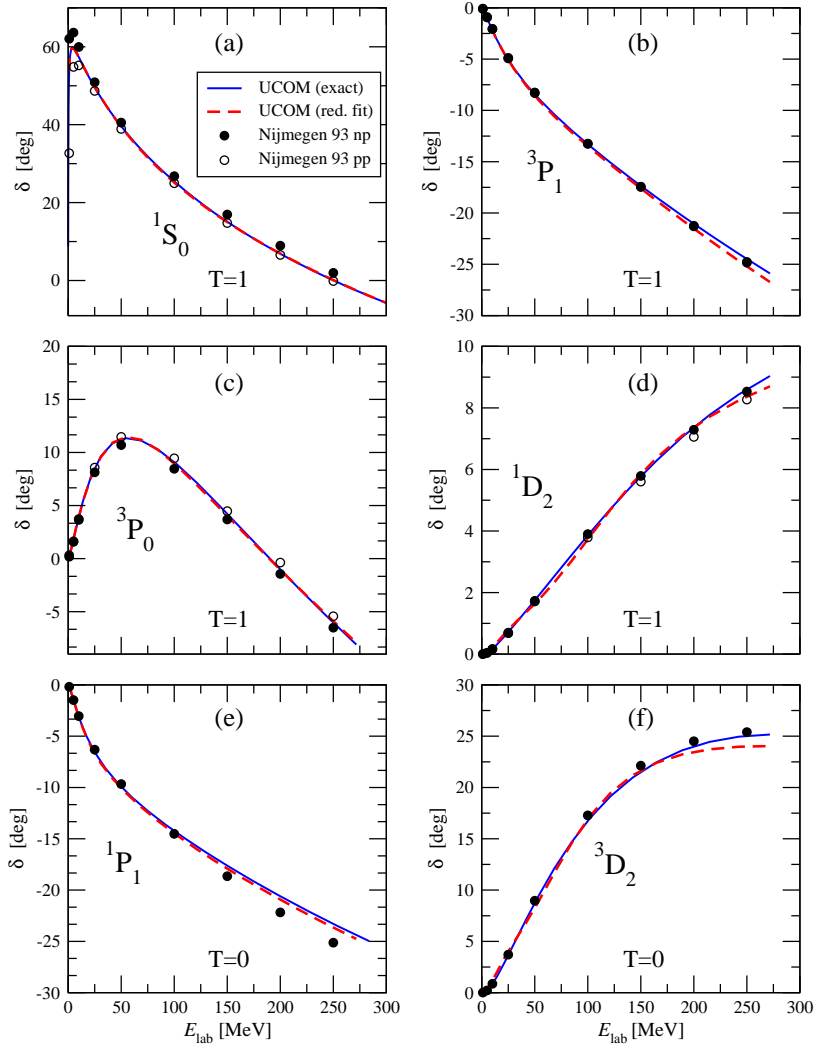}
\caption{\label{fig:phs1}(Color online) NN phase shifts for total angular momentum $J=0$, $1$ and $2$ calculated with the exact matrix elements of the UCOM(0.04) transformed Argonne potential (blue solid line) and the UCOM(0.04) fit with the reduced set of operators (red dashed line). The dots indicate the results of the 1993 Nijmegen partial wave analysis \cite{PhysRevC.48.792}. The phase shifts of the original Argonne potential are identical with those of the UCOM transformed Argonne potential.}
\end{figure*}

\begin{figure*}[ht!]
\centering
\includegraphics[width=0.6\textwidth]{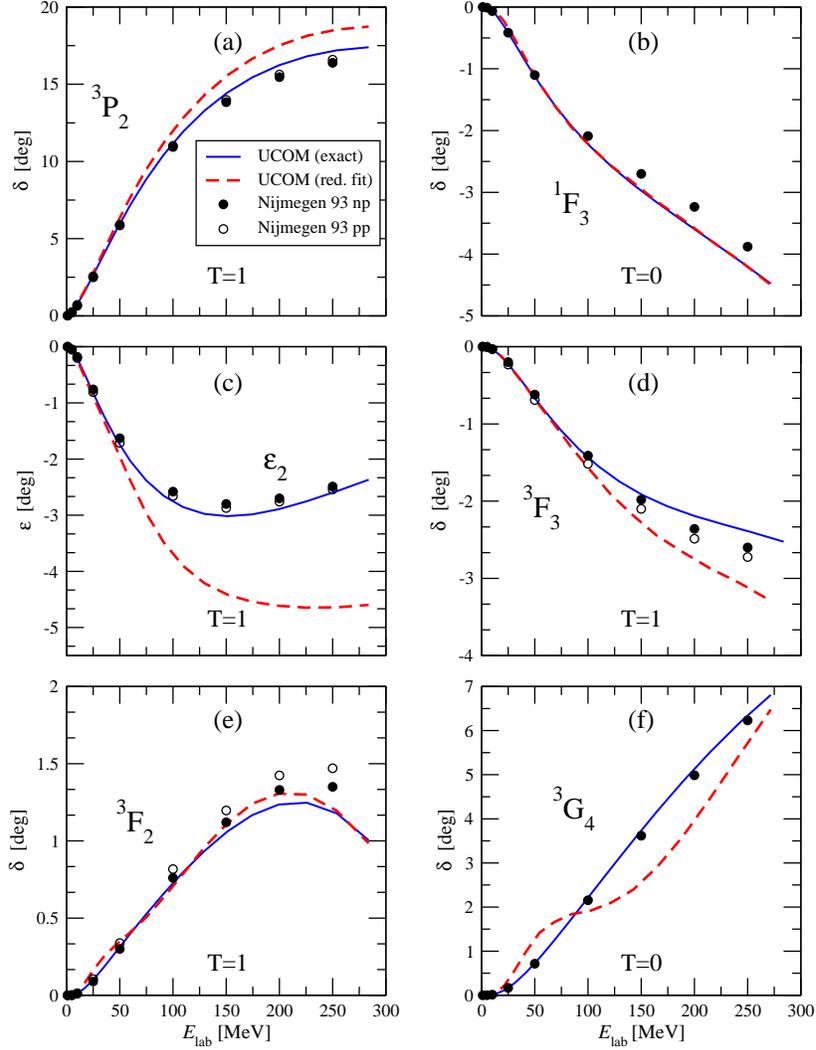}
\caption{\label{fig:phs2}(Color online) Same as Fig.~\ref{fig:phs1}, but for $J=2$ and $J=3$. Note that the absolute scales are smaller than in Fig.~\ref{fig:phs1}.}\end{figure*}

Figs.~\ref{fig:neglected} - \ref{fig:phs2} show the phase shifts calculated with the exact UCOM(0.04) matrix elements and those of the reduced UCOM(0.04) fit. The results for the deuteron can be found in Tab.~\ref{tab:deuteronres}. 
\begin{table}[h!]
\centering
\begin{ruledtabular}
\begin{tabular}{l|*{3}{.}}
\multicolumn{1}{c|}{$^2\rm{H}$}& \multicolumn{1}{c}{$E_B\,[\mbox{MeV}]$} & \multicolumn{1}{c}{$\mu\, [\mu_N]$} & \multicolumn{1}{c}{$Q\, [e\,\mbox{fm}^2]$}  \\ \hline
UCOM (exact) & 2.23 & 0.847 & 0.270 \\
UCOM (red. fit)& 2.23 & 0.847 & 0.271 \\ 
Experiment & 2.2246 & 0.8574 & 0.2860 
\end{tabular}
\end{ruledtabular}
\caption{\label{tab:deuteronres} Binding energy $E_B$, magnetic dipole moment $\mu$ and electric quadrupole moment $Q$ of the deuteron for the exact UCOM(0.04) transformed Argonne matrix elements and the reduced UCOM(0.04) fit compared with experimental data. The results for $\mu$ and $Q$ have been obtained with correlated operators. Experimental data from Ref.~\cite{Dumbrajs1983277}.}
\end{table}

The phase shifts show very good agreement up to the D-wave. Deviations occur for $L=3$ and above, especially at higher laboratory energies. These deviations are in general rather small compared to the absolute value of the phase shifts. The deuteron properties, which are sensitive only to the $^3\rm{S}_1$ and $^3\rm{D}_1$ channels, are described very well by the reduced UCOM(0.04) fit and the exact UCOM(0.04) matrix elements. 

This shows that both interactions have the same features at low angular momenta: phase shifts and deuteron properties are in good agreement. 

At higher angular momenta $L$ the reduced UCOM potential reproduces the phase shifts with larger relative error, but the obtained absolute values differ also by not more than 2 degrees (see Fig.~\ref{fig:phs2}).

\subsection{Few-nucleon systems with NCSM} 

We employ the {\it ManyEff} code by Petr Navr$\acute{\rm{a}}$til \cite{PhysRevC.61.044001} for nuclei with $A=3$ and $4$. The calculations for larger mass numbers are performed with the {\it Antoine} NCSM code \cite{shellmodeltechniques,PhysRevC.64.051301}.

The binding energies and other properties of selected light nuclei are calculated using the exact and the reduced UCOM matrix elements. The model space sizes (up to $N_{\mathrm{max}}$ excitations above the $0\hbar\Omega$ configuration) are $N_{\mathrm{max}}=40$ for $^3\mathrm{H}$ and $^3\mathrm{He}$, $N_{\mathrm{max}}=16$ for $^4\mathrm{He}$, $N_{\mathrm{max}}=12$ for $^6\mathrm{He}$ and $^6\mathrm{Li}$, and $N_{\mathrm{max}}=10$ for $^7\mathrm{Li}$. For nuclei with $A\geq4$ the energies obtained with the NCSM do not converge perfectly in the used model space sizes up to $N_{\mathrm{max}}$. The converged energies are obtained from a simple exponential extrapolation to infinite model space size. Because the NCSM calculation is variational we perform the extrapolation in the model space with the oscillator parameter $\hbar\Omega$  yielding the lowest energy at $N_{\mathrm{max}}$. They are $\hbar\Omega=16\,\mathrm{MeV}$ for $^4\mathrm{He}$, $\hbar\Omega=20\,\mathrm{MeV}$ for $^6\mathrm{He}$ and $^6\mathrm{Li}$, and $\hbar\Omega=24\,\mathrm{MeV}$ for $^7\mathrm{Li}$. For the nuclei discussed here, the results are well-converged and we do not need to use more elaborate extrapolation schemes \cite{PhysRevC.86.031301}. 

\begin{table*}[h]
\centering
\begin{ruledtabular}
\begin{tabular}{l  | *{6}{.}}
 &\multicolumn{1}{c}{$^3\mbox{H}$} & \multicolumn{1}{c}{$^3\mbox{He}$} & \multicolumn{1}{c}{$^4\mbox{He}$} & \multicolumn{1}{c}{$^6\mbox{He}$} & \multicolumn{1}{c}{$^6\mbox{Li}$} & \multicolumn{1}{c}{$^7\mbox{Li}$} \\  \hline
UCOM (exact) & 8.38 & 7.67 & 28.53 & 28.4 & 31.5 & 38.6 \\
UCOM (red. fit) & 8.37 & 7.66 & 28.52 & 28.5 & 31.6 & 38.7 \\ 
Experiment  & 8.482 & 7.718 & 28.296 & 29.269 & 31.995 & 39.245
\end{tabular}
\end{ruledtabular}
\caption{\label{tab:lightnuclei} Binding energies (in $\rm{MeV}$) of some light nuclei calculated in the NCSM with the exact matrix elements of the UCOM(0.04) transformed Argonne potential (UCOM (exact)) and the UCOM(0.04) fit with the reduced set of operators (UCOM (red. fit)). The results for the $^4\mbox{He}$, $^6\mbox{He}$, $^6\mbox{Li}$ and $^7\mbox{Li}$ binding energies are obtained by an extrapolation to infinite model space size.}
\end{table*}
\begin{table*}[h!]
\centering
\begin{ruledtabular}
\begin{tabular}{l  | c *{4}{.}}
 &\multicolumn{1}{c}{Nucl.}& \multicolumn{1}{c}{$E_B\,\rm{[MeV]}$}& \multicolumn{1}{c}{$R_p\,\rm{[fm]}$} & \multicolumn{1}{c}{$\mu\,[\mu_N]$} & \multicolumn{1}{c}{$Q\,[e\,\rm{fm}^2]$}  \\  \hline
UCOM (exact) & & 31.5 & 2.1 & 0.843 & -0.04  \\ 
UCOM (red. fit) &$^6\rm{Li}$ & 31.6 &  2.1 & 0.842 & -0.03  \\
Experiment & & 31.995 & 2.41(3) & 0.8220 & -0.0818(17)  \\ \hline
UCOM (exact) & & 38.6 & 2.0 & 2.988 & -2.6  \\ 
UCOM (red. fit) &$^7\rm{Li}$& 38.7 & 2.0 & 2.987 & -2.6  \\
Experiment & & 39.245 & 2.26(2) & 3.2564 & -4.06(8)
\end{tabular}
\end{ruledtabular}
\caption{\label{tab:lightnuclei2}Properties of $^6\rm{Li}$ and $^7\rm{Li}$. Binding energy $E_B$, point-proton radius $R_p$, magnetic dipole moment $\mu$ and electric quadrupole moment $Q$, calculated in the NCSM with the exact UCOM(0.04) matrix elements (UCOM (exact)) and the reduced UCOM(0.04) fit (UCOM (red. fit)). The energies are obtained from an extrapolation to infinite model space size. The other properties are calculated in a model space size of $12\hbar\Omega$ for $^6\rm{Li}$ and $10\hbar\Omega$ for $^7\rm{Li}$ with an oscillator frequency of $\hbar\Omega=24\,\rm{MeV}$ and by means of bare operators. Experimental data from Ref.~\cite{Tilley20023} and \cite{PhysRevLett.96.033002}.}
\end{table*}

The (extrapolated) binding energies are shown in Tab.~\ref{tab:lightnuclei}. The binding energies obtained with the reduced set of operators lie within a range of $10$ to $100\,\mbox{keV}$ around those calculated with the exact UCOM(0.04) matrix elements, which corresponds to relative deviations of less than $0.5\%$

\begin{figure*}[h!]
\begin{minipage}{0.45\textwidth}
\centering
\includegraphics[width=\textwidth]{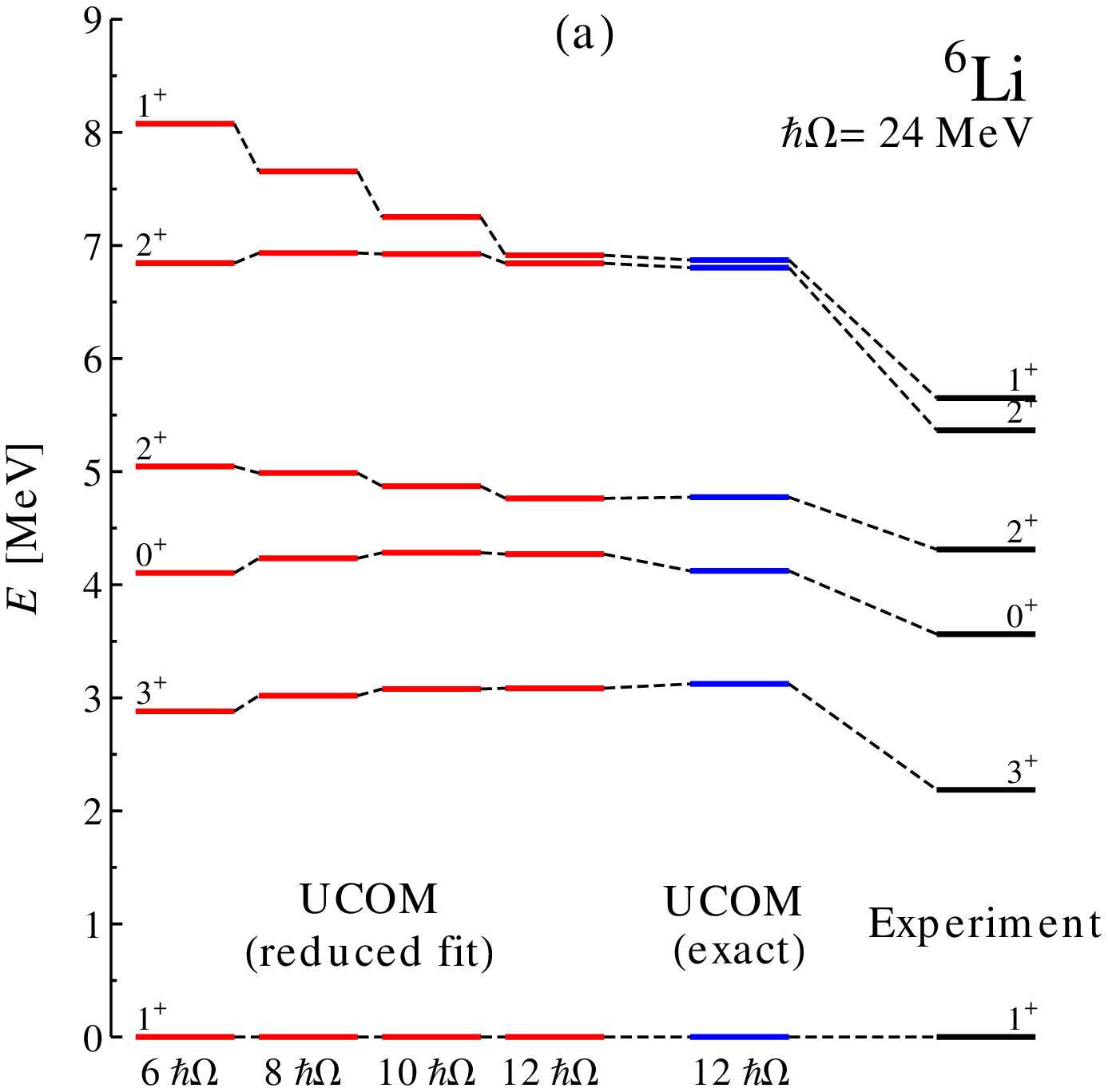}
\end{minipage}
\begin{minipage}{0.45\textwidth}
\centering
\includegraphics[width=\textwidth]{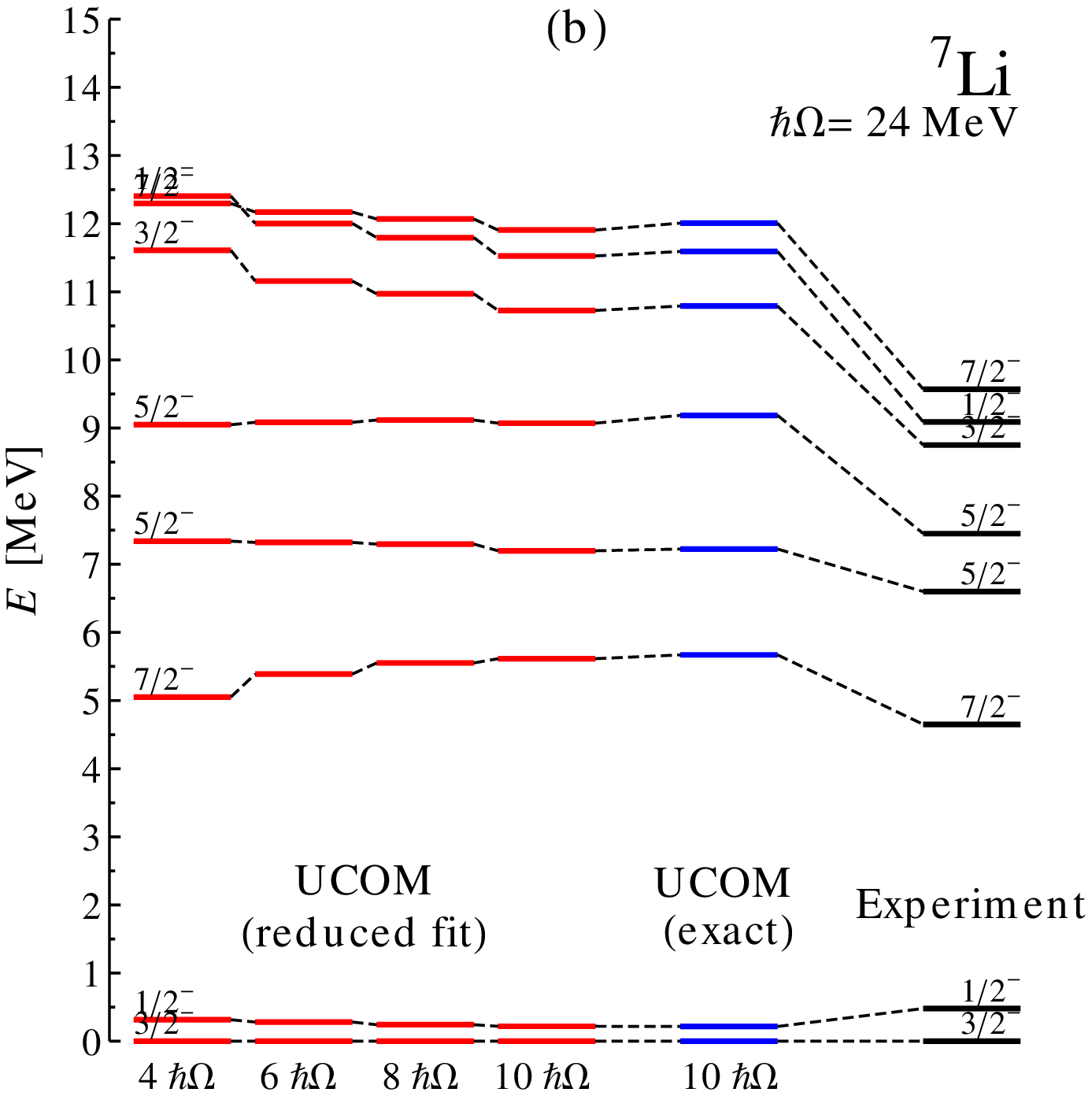}
\end{minipage}
\caption{\label{fig:spec}(Color online) Spectra of (a) $^6\rm{Li}$ and (b) $^7\rm{Li}$. The red lines show the results of a NCSM calculation with the matrix elements of the reduced UCOM(0.04) fit for different model spaces up to $12\,\hbar\Omega$ for $^6\rm{Li}$ and up to $10\,\hbar\Omega$ for $^7\rm{Li}$ in comparison with the results of the exact UCOM(0.04) matrix elements (blue lines) in a $12\,\hbar\Omega$ and $10\,\hbar\Omega$ model space and the experiment \cite{Tilley20023}.
}
\end{figure*}

Good agreement can also be seen for calculated ground state properties in Tab.~\ref{tab:lightnuclei2}, which shows results for the point proton radius, the magnetic dipole moment and the electric quadrupole moment of $^6\rm{Li}$ and $^7\rm{Li}$. These results still depend on the size of the model space and no extrapolations are used. However, we observe good agreement between the results from the exact and reduced UCOM potential. The nuclear spectra of $^6\mbox{Li}$ and $^7\mbox{Li}$, calculated with the reduced UCOM(0.04) and the exact UCOM(0.04) matrix elements, are shown in Fig.~\ref{fig:spec}. By comparing the $12\hbar\Omega$ results of the reduced UCOM potential with the exact UCOM result for $^6\mbox{Li}$ and the $10\hbar\Omega$ results for $^7\mbox{Li}$, we see that both interactions clearly lead to almost identical energy spectra.\\

\subsection{Few-nucleon systems with fermionic molecular dynamics (FMD)\label{sec:FMD}}

In FMD \cite{RevModPhys.72.655,Neff200869,PhysRevLett.106.042502,PhysRevLett.108.142501} many-body basis states $\ket{Q}$ are given by antisymmetrized product states
\begin{equation}
  \ket{Q} = \op{\mathcal{A}}\big(\ket{q_1}\otimes\ket{q_2}\otimes\cdots\otimes\ket{q_A}\big),
  \label{eq:FMDQ}
\end{equation}
where the single particle states
\begin{equation}
\ket{q}=\sum_{k=1}^nc_{k}\ket{a_{k}\,\vec{b}_{k},\chi_{k};\xi}\label{eq:FMDstate}
\end{equation}
are superpositions of Gaussian wave packets
\begin{equation}
  \braket{\vec{r}}{a,\vec{b}} = \exp \left\{- \frac{(\vec{r}-\vec{b})^2}{2a} \right\}.
  \label{eq:FMDgauss}
\end{equation}
$\chi$ denotes a two-component spinor and $\xi$ is the isospin of the nucleon. Because of the flexibility of the wave packet basis, not only shell-model-like states, but also exotic structures like clusters and halos can be described with a numerically feasible effort.

Since the intrinsic FMD basis states $\ket{Q}$ may break the symmetries of the Hamiltonian under reflection, rotation and translation, these symmetries have to be restored by projection on parity, angular momentum, and total momentum zero:
\begin{equation}
 \ket{Q; J^\pi MK; \vec{P}=0} = \op{P}^\pi \op{P}^J_{MK} \op{P}^{\vec{P}=0} \ket{Q}.
 \label{eq:FMDQproj}
\end{equation}

To describe the ground state and the excited states of a nucleus, a set of basis states $\ket{Q^{(i)}}$ will be used. The basis states are determined by the parameters $a$, $\vec{b}$ and $\chi$ of the single-particle states which are obtained by variation. In the simplest approach the variation is performed for the energy expectation value of the intrinsic state Eq.~(\ref{eq:FMDQ}) and the state is only projected after variation. This approach is essentially a mean-field calculation with restoration of symmetries. To improve the description of correlations -- both long-range correlations like clustering and short-range correlations as induced by the tensor force -- the variation should be done for the projected state Eq.~(\ref{eq:FMDQproj}). This variation after projection (VAP) is performed for all spins of the nucleus independently. Additional many-body basis states can be obtained by using constraints like radius or quadrupole deformation. The final results are obtained by diagonalizing the Hamiltonian in the set of projected non-orthogonal many-body basis states $\ket{Q^{(i)}; J^\pi MK; \vec{P}=0}$.

Because the parameters of the FMD states change in each step of the variation procedure, the matrix elements can not be computed in advance, and it is essential to have an efficient way to calculate the matrix elements. Furthermore, the non-orthogonality of the basis and the need for gradients of the matrix elements make it necessary to have analytical expressions for the matrix elements. If the NN potential is given in operator representation, for example as a sum of operators $\op{\mathcal{O}}_P$ with the corresponding radial functions $V^P_{ST}(\op{r})$:
\begin{equation}
\op{V}_{\mathrm{NN}}=\sum_P\sum_{ST}\frac{1}{2}\left[\op{\mathcal{O}}_P{V}^P_{ST}(\op{r})+{V}^P_{ST}(\op{r})\op{\mathcal{O}}_P\right],
\end{equation}
analytical expressions can be found if the radial functions $V^P_{ST}(\op{r})$ are represented by sums of Gaussians
\begin{subequations}
\begin{equation}
V^P_{ST}(r)=\sum_{\mu}\gamma^P_{ST,\mu}G_{\mu}(r),
\label{eq:parrad}
\end{equation}
with
\begin{eqnarray*}
G_{\mu}(r)=\mbox{exp}\left\{-\frac{r^2}{2\kappa_{\mu}}\right\}.
\end{eqnarray*}
In case of the tensor operators $\op{S}_{12}$, $\bar{S}_{12}(\vec{\op{p}}_{\Omega},\vec{\op{p}}_{\Omega})$ and $S_{12}(\vec{\op{r}},\vec{\op{p}}_{\Omega})$ the representation
\begin{eqnarray}
V^{T/Tpp}_{ST}(r)&=&\sum_{\mu}\gamma^{T/Tpp}_{ST,\mu}\,\,r^2\cdot G_{\mu}(r) \label{eq:parrad2}\\
V^{Trp}_{ST}(r)&=&\sum_{\mu}\gamma^{Trp}_{ST,\mu}\,\,r^3\cdot G_{\mu}(r)
\label{eq:parrad3}
\end{eqnarray}
\label{eq:FMDrad}
\end{subequations}
is used for convenience.

The expressions for the FMD matrix elements for the full and reduced operator representation of the UCOM potential (Eqs.~(\ref{eq:ucompotential}) and (\ref{eq:reducompotential})) are given in Appendix~\ref{sec:fmdme}.

\begin{table}
\begin{ruledtabular}
 \begin{tabular}{lc|..|..}
  & & \multicolumn{2}{c|}{UCOM(0.04)} & \multicolumn{2}{c}{UCOM(0.20)} \\
  & & \multicolumn{1}{c}{full fit} & \multicolumn{1}{c|}{red. fit} & \multicolumn{1}{c}{full fit} & \multicolumn{1}{c}{red. fit} \\
  \hline
  FMD(V) & $E_B$ &  19.14 &  19.14 &  25.43 &  25.43 \\
      & $T$   &        &  50.60 &        &  48.53 \\
      & $V_c$ &        & -70.59 &        & -74.80 \\
      & $V_{\mathit{Coul}}$ &    &  0.86  &        &0.86 \\
  \hline
  FMD(VAP) & $E_B$ & 26.97 &  26.97 &  27.91 &  27.90 \\
      & $T$   &        &  58.36 &        &  50.34 \\
      & $V_c$ &        & -72.54 &        & -75.17 \\
      & $V_\mathit{ls}$ & & 0.40 &       &   0.08 \\
      & $V_T$ &        & -14.06 &        &  -3.98 \\
      & $V_{\mathit{Coul}}$ &    &  0.83  &        &0.82 \\
  \hline
  NCSM (extrapolated) & $E_B$ & 28.53 &  28.52 &  28.57 &  28.58
 \end{tabular}
\end{ruledtabular}
\caption{FMD results for $^4$He obtained by simple variation (V) and variation after projection (VAP) using UCOM interactions obtained from Argonne~V18 with both full and reduced sets of operators and for flow parameters $\alpha=0.04\,\mathrm{fm}^4$ and $\alpha=0.2\,\mathrm{fm}^4$. The contributions from the kinetic energy, central, spin-orbit, and tensor contributions of the transformed interactions are given for the reduced interactions. The exact binding energies as obtained in NCSM calculations are given for comparison. All energies are in MeV.}
\label{tab:He4fmd}
\end{table}

To concentrate on the properties of the interaction, we present here only a basic FMD calculation for the $^4$He ground state. The wave function is a single FMD four-body state where a superposition of two Gaussians (Eq.~(\ref{eq:FMDstate})) is used for each single-particle state. To illustrate the role of correlations, we compare in Tab.~\ref{tab:He4fmd} the results obtained by simple variation (V) with the more sophisticated variation after projection calculation (VAP). In the mean-field like approach (V) the wave function is spin-saturated and thus there is no contribution from the spin-orbit and tensor components of the UCOM interaction. The binding energy obtained for the UCOM(0.04) interaction with flow parameter $\alpha=0.04\,\mathrm{fm}^4$ is more than 9~MeV lower than the exact result as obtained from the NCSM calculation. The FMD(VAP) calculation, on the other hand, underestimates the binding energy only by 1.5~MeV. This is explained by the tensor correlations in the wave function. In the VAP calculation the tensor components of the UCOM interaction contribute with $-14.06$~MeV to the potential energy. The correlations in the wave functions also lead to an increase in the kinetic energy so that the total binding energy increases only by 7.83~MeV.

\begin{figure*}[ht!]
\centering
\includegraphics[width=0.85\textwidth]{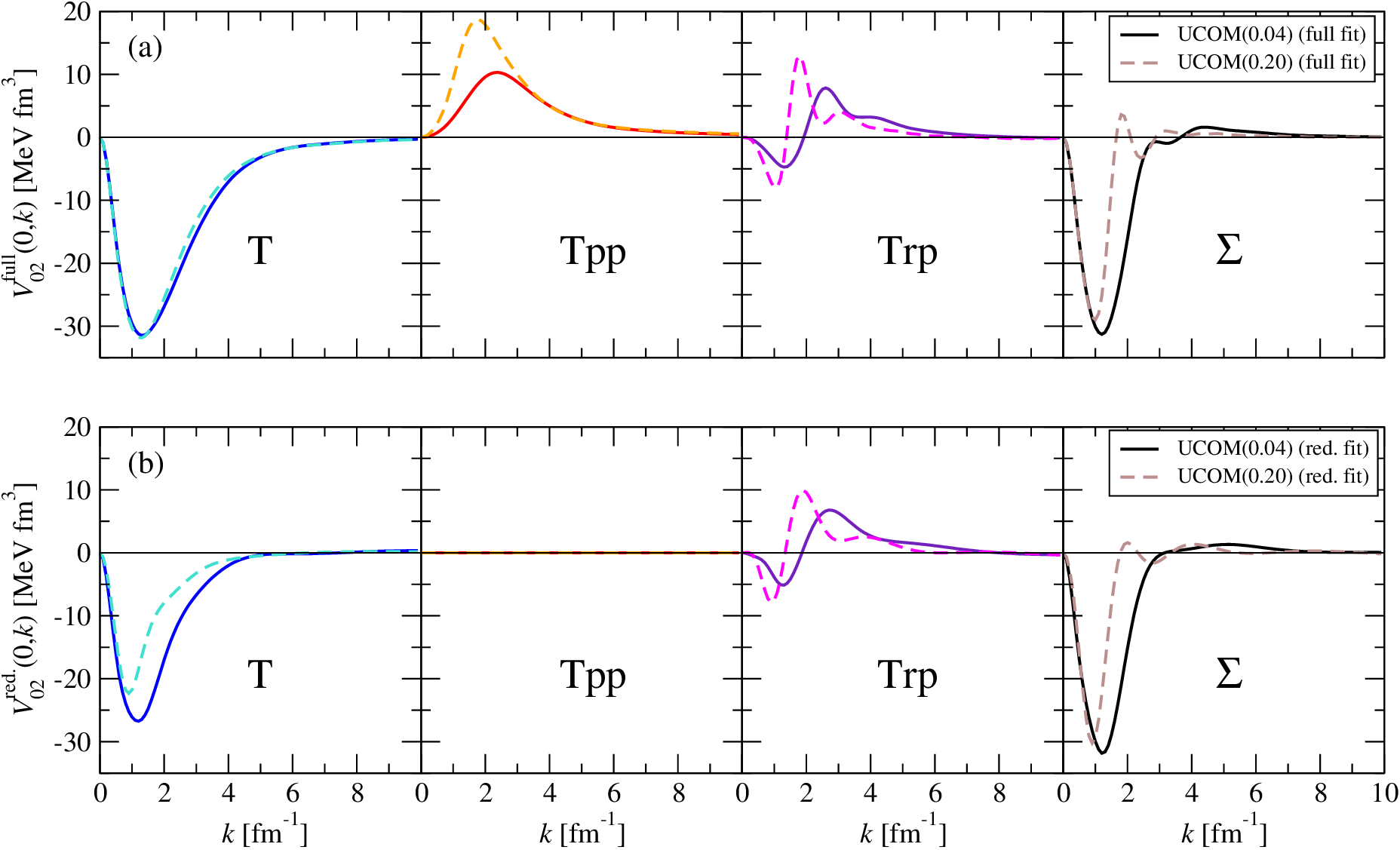}
\caption{\label{fig:tensormes}(Color online) Off-diagonal matrix elements $V_{02}(0,k):=\bra{0(01)1;0}\op{V}\ket{k(21)1;0}$ of the UCOM(0.04) and UCOM(0.20) fit with (a) the full set of operators and (b) the reduced set of operators. From left to right: Contributions from the $\op{S}_{12}$ tensor term, the $\bar{S}_{12}(\vec{\op{p}}_{\Omega},\vec{\op{p}}_{\Omega})$ tensor term, the momentum dependent tensor $S_{12}(\vec{\op{r}},\vec{\op{p}_{\Omega}})$ and the sum of these contributions.}
\end{figure*}

Although the tensor force in the UCOM interaction is much weaker than in the original Argonne interaction, the correlations induced by the tensor components in the UCOM interaction are still difficult to describe in the FMD approach in case of $^4\rm{He}$, and it becomes even more difficult for heavier nuclei. It is therefore advantageous to use a softer interaction as can be obtained with a larger flow parameter like $\alpha=0.2\,\mathrm{fm}^4$, UCOM(0.20). For this interaction a larger part of the original tensor force has been renormalized and the remaining tensor force in the UCOM interaction is significantly weaker. This is illustrated in Fig.~\ref{fig:tensormes} by means of the off-diagonal momentum space matrix elements of the $^3\rm{S}_1$-$^3\rm{D}_1$ channel, which contains only tensor contributions. Looking at the UCOM potential with the full set of operators, the UCOM transformation creates with increasing flow parameter $\alpha$ stronger contributions of the $\bar{S}_{12}(\vec{\op{p}}_{\Omega},\vec{\op{p}}_{\Omega})$ and $S_{12}(\vec{\op{r}},\vec{\op{p}_{\Omega}})$ operators that cancel the contributions of the tensor operator $\op{S}_{12}$ at larger momenta. For the reduced set of operators, the contributions of $\bar{S}_{12}(\vec{\op{p}}_{\Omega},\vec{\op{p}}_{\Omega})$ are absorbed in the $\op{S}_{12}$ term, which therefore shows a stronger $\alpha$ dependence. The reduced UCOM fit shows with increasing $\alpha$ the same reduction of the high momentum tensor components as the full UCOM fit. This reduced tensor force also leads to reduced tensor correlations in the exact NCSM wave function and the binding energy with the single FMD basis state gets within 0.7~MeV of the NCSM result for the UCOM(0.20) interactions.

\section{Conclusions}
We have derived a reduced UCOM potential, which contains less operators than the exact UCOM transformed Argonne potential.

In calculations of two-nucleon systems, discussed in Sec.~\ref{sec:twofewsystems}, reduced UCOM and the exact UCOM transformed Argonne potential show the same results for the deuteron and the phase shifts up to the D-wave and slight deviations in the phase shifts with higher angular momenta. The latter can be traced back to the fact that the reduced set of operators cannot perfectly describe the exact UCOM matrix elements in all partial waves and that the fitting method puts more emphasis on low angular momentum matrix elements to be reproduced as accurately as possible. The small deviations for higher angular momenta, however, have no effect on the results for the few-nucleon systems investigated by means of NCSM and FMD calculations. The reduced UCOM potential yields the same results as the exact UCOM transformed Argonne potential for a wide range of different physical properties of light nuclei, such as energies, energy level spectra, radii, magnetic dipole and electric quadrupole moments. 

From these results one can conclude that the reduced UCOM potential contains all the important features of the exact UCOM transformed Argonne potential that are relevant to describe the light nuclear systems discussed. At the same time it contains a smaller set of operators. Due to this reduced set of operators, the reduced UCOM potential allows to perform FMD calculations for light nuclei with a reduced computational effort and without loosing precision at the same time. 

Furthermore, the structure of the reduced UCOM potential allows to draw conclusions on the importance of additional operators in soft effective interactions beyond those already used in the Argonne potential Eq.~\eqref{eq:argonnepotential}. Although many new operators are created by the UCOM transformation, one succeeds in describing the important features of a realistic effective interaction by the set of operators already present in the bare Argonne potential plus the two momentum dependent operators given in Eq.~\eqref{eq:ucompotential}. This shows that especially the momentum dependent operators, which replace the short-range repulsion and short-range tensor force,  play an essential role for the potential and therefore have to be included in the reduced set of operators. The influence of the other operators created by the UCOM transformation is either small or can be absorbed by already included operators without altering two- and few-nucleon properties significantly, so that these operators do not have to be 
considered explicitly in the reduced set of operators.

In future studies we plan to derive operator representations for other effective realistic NN interactions, for example potentials based on SRG transformations. The question of how local and non-local components of these interactions can be disentangled has recently been discussed in  \cite{PhysRevC.86.014003}. In contrast to the UCOM transformation, which yields only quadratic momentum terms, the SRG creates already for local initial interactions a more complicated, non-polynomial momentum dependence. Consequently, the ansatz for an operator representation of SRG transformed potentials must contain a more complex momentum dependence to be flexible enough to describe the nonlocal structure of these potentials. 

\section{Acknowledegments}
This work was supported by the Helmholtz Alliance Program of the Helmholtz Association, contract HA216/EMMI "Extremes of Density and Temperature: Cosmic Matter in the Laboratory". H. Hergert acknowledges support by the the National Science Foundation under Grant No. PHY-10002478 and the NUCLEI SciDAC Collaboration under the U.S. Department of Energy Grant No. DE-SC0008533.

\appendix

\section{UCOM matrix elements\label{sec:appendixUCOMmes}}
The UCOM transformation can be performed exactly in a basis with good angular momentum and spin quantum numbers. In this appendix, we show the matrix elements for the partial wave basis in momentum space $\ket{k(LS)J;T}$, where  $k$ is the relative momentum, $L$ the relative angular momentum, $S$ the total spin, $J$ the total angular momentum and $T$ the total isospin. With the definitions
\begin{subequations}
\begin{eqnarray}
\widehat{\Theta}_{SJT}(r)&=&3\sqrt{J(J+1)}\vartheta_T(R_{+,ST}(r)) \\
W_{ST}(r)&=&\frac{7R''_{+,ST}(r)^2}{4R'_{+,ST}(r)^4}-\frac{R'''_{+,ST}(r)}{2R'_{+,ST}(r)^3} \\
\frac{1}{2\mu_{r,ST}(r)}&=&\frac{1}{2\mu}\Big(R'_{+,ST}(r)^{-2}-1\Big),
\end{eqnarray}
\end{subequations}where $R'_{+,ST}(r)=\frac{\mathrm{d}}{\mathrm{d}r}R_{+,ST}(r)$, one finds for the momentum matrix elements of the correlated Argonne potential with the operators $\op{\mathcal{O}}$ in Eq.~(\ref{eq:operators}) and the matrix elements of the correlated kinetic energy \cite{PhysRevC.72.034002}:
\begin{widetext}
\begin{subequations}
\begin{eqnarray}
&&\bra{k(JS)J;T} \op{C}_r^{\dag}\op{C}_{\Omega}^{\dag}v(\op{r}) \op{\mathcal{O}}\op{C}_{\Omega}\op{C}_r\ket{k'(JS)J;T} = \frac{2}{\pi}\int_{0}^{\infty} \mbox{d}rr^2 \, j_J(kr)\widehat{v}(r)j_J(k'r) \bra{(JS)J;T}\op{\mathcal{O}}\ket{(JS)J;T}\label{eq:vme1} \\ \nonumber
\\
&&\bra{k(J\mp 1 S)J;T} \op{C}_r^{\dag}\op{C}_{\Omega}^{\dag}v(\op{r})\op{\mathcal{O}}\op{C}_{\Omega}\op{C}_r\ket{k'(J\pm1S)J;T}\nonumber \\
&&\quad= 
\frac{2}{\pi}\int_0^{\infty}\!\!\!\!\!\! \mbox{d}rr^2\, j_{J\mp1}(kr)\widehat{v}(r)j_{J\pm1}(k'r) \cdot 
\Big[\bra{(J\mp1S)J;T}\op{\mathcal{O}}\ket{(J\pm1S)J;T}\;\left(\cos\left(\widehat{\Theta}_{SJT}(r)\right)\right)^2 \nonumber \\
&&\qquad-\bra{(J\pm1S)J;T}\op{\mathcal{O}}\ket{(J\mp1S)J;T}\;\left(\sin\left(\widehat{\Theta}_{SJT}(r)\right)\right)^2 \mp\bra{(J\mp1S)J;T}\op{\mathcal{O}}\ket{(J\mp1S)J;T}\frac{1}{2}\sin \left(2\widehat{\Theta}_{SJT}(r)\right) \nonumber \\
&&\qquad\pm\bra{(J\pm1S)J;T}\op{\mathcal{O}}\ket{(J\pm1S)J;T}\frac{1}{2}\sin \left(2\widehat{\Theta}_{SJT}(r)\right)\Big]\\ \nonumber
\\
&&\bra{k(J\mp 1 S)J;T} \op{C}_r^{\dag}\op{C}_{\Omega}^{\dag}v(\op{r}) \op{\mathcal{O}}\op{C}_{\Omega}\op{C}_r\ket{k'(J\mp1S)J;T} \nonumber \\
&&\quad=\frac{2}{\pi}\int_0^{\infty}\!\!\!\!\!\! \mbox{d}rr^2\, j_{J\mp1}(kr)\widehat{v}(r)j_{J\mp1}(k'r)\Big[\bra{(J\mp1S)J;T}\op{\mathcal{O}}\ket{(J\mp1S)J;T}\;\left(\cos\left(\widehat{\Theta}_{SJT}(r)\right)\right)^2 \nonumber \\
&&\qquad+\bra{(J\pm1S)J;T}\op{\mathcal{O}}\ket{(J\pm1S)J;T}\;\left(\sin\left(\widehat{\Theta}_{SJT}(r)\right)\right)^2 \pm\bra{(J\mp1S)J;T}\op{\mathcal{O}}\ket{(J\pm1S)J;T}\sin \left(2\widehat{\Theta}_{SJT}(r)\right)\Big]\nonumber\\
\label{eq:vme3}\\
&&\bra{k(JS)J;T} \op{C}_r^{\dag}\op{C}_{\Omega}^{\dag}\op{T}_r\op{C}_{\Omega}\op{C}_r\ket{k'(JS)J;T}\nonumber \\
&&\quad= \frac{2}{\pi}\int_0^{\infty}\!\!\!\!\!\! \mbox{d}rr^2 \Bigg\{ j_J(kr)j_J(k'r)\cdot \Bigg[\frac{W_{ST}(r)}{2\widehat{\mu}_{r,ST}(r)}+\frac{\widehat{\mu}_{r,ST}'(r)}{2\widehat{\mu}_{r,ST}(r)^2}\frac{{R}''_{+,ST}(r)}{\left(R'_{+,ST}(r)\right)^2}\Bigg] \nonumber \\
&&\qquad-\frac{1}{2}\Big[k^2j''_J(kr)j_J(k'r)+ j_J(kr)k'^2j''_J(k'r)\Big]\cdot\frac{1}{2\widehat{\mu}_{r,ST}(r)\left(R_{+,ST}(r)\right)^2}\Bigg\}\\
 \nonumber \\
&&\bra{k(J\mp1S)J;T} \op{C}_r^{\dag}\op{C}_{\Omega}^{\dag}\op{T}_r\op{C}_{\Omega}\op{C}_r\ket{k'(J\pm1S)J;T}\nonumber \\
&&\quad=\pm \frac{2}{\pi}\int_0^{\infty}\!\!\!\!\!\! \mbox{d}rr^2 \Big[j_{J\mp1}(kr)k'j'_{J\pm1}(k'r) -k j'_{J\mp1}(kr)j_{J\pm1}(k'r)\Big]\cdot\frac{\widehat{\Theta}_{SJT}'(r)}{2\widehat{\mu}_{r,ST}(r)R'_{+,ST}(r)^2}\label{eq:tme} \\
\nonumber
\\
&&\bra{k(J\mp1S)J;T} \op{C}_r^{\dag}\op{C}_{\Omega}^{\dag}\op{T}_r\op{C}_{\Omega}\op{C}_r\ket{k'(J\mp1S)J;T}\nonumber \\
&&\quad=  \frac{2}{\pi}\int_0^{\infty}\!\!\!\!\!\! \mbox{d}rr^2 \Bigg\{ j_{J\mp1}(kr)j_{J\mp1}(k'r)\cdot\Bigg[\frac{W_{ST}(r)}{2\widehat{\mu}_{r,ST}(r)}+\frac{\widehat{\Theta}_{SJT}'(r)^2}{2\widehat{\mu}_{r,ST}(r)}+\frac{\widehat{\mu}'_{r,ST}(r)}{2\widehat{\mu}_{r,ST}(r)^2}\frac{R_{+,ST}''(r)}{R_{+,ST}'(r)^2}\Bigg] \nonumber \\
&&\qquad-\frac{1}{2}\Big[j_{J\mp1}(kr)k'^2j''_{J\mp1}(k'r) + k^2j''_{J\mp1}(kr)j_{J\mp1}(k'r)\Big]\cdot\frac{1}{2\widehat{\mu}_{r,ST}(r)R'_{+,ST}(r)^2}\Bigg\}.
\end{eqnarray}
\label{eq:ucommat}
\end{subequations}
\end{widetext}
$j'_{L}(x)$ stands for the derivative of the spherical Bessel function $j_{L}(x)$. The matrix elements for the angular part $\op{T}_{\Omega}=\frac{1}{2\mu}\frac{\vec{\op{L}}^{\,2}}{\op{r}^2}$ can be obtained from Eqs.~(\ref{eq:vme1}) to (\ref{eq:vme3}) by setting $v(r)=\frac{1}{2\mu r^2}$ and $\op{\mathcal{O}}=\vec{\op{L}}^{\,2}$.

\section{Fit of the operator representation}

\subsection{Radial integrals\label{sec:radint}}
To find an analytic expression for the matrix elements Eq.~\eqref{eq:ansatzme} of the reduced UCOM potential one has to calculate the integral
\begin{align}
I_{\kappa}\left(k,k',L,L';n_P\right):=&\nonumber \\
\int_0^{\infty}drr^2 j_L(kr)&\,r^{n_P}\mbox{exp}\left\{-\frac{r^2}{2\kappa_{\mu}}\right\}j_{L'}(k'r)
\end{align}
occuring in Eq.~\eqref{eq:ansatzme} and \eqref{eq:ansatzmemom}. The explicit expressions for $n_P=0$ and $L=L'\leq4$ are:
\begin{widetext}
\begin{subequations}
\begin{align}
I_{\kappa}\left(k,k',0,0;0\right)&=&\int_{0}^{\infty} \mbox{d}r\, r^2 j_{0}(kr)&\,e^{-\frac{r^2}{2\kappa}} j_{0}(k'r)
=\frac{\sqrt{2\pi}}{4}\frac{1}{k k' \kappa^{-1/2}}\Big[e^{-\frac{\kappa}{2}(k-k')^2}-e^{-\frac{\kappa}{2}(k+k')^2}\Big] \\[0.3cm] 
I_{\kappa}\left(k,k',1,1;0\right)&=&\int_{0}^{\infty} \mbox{d}r\, r^2 j_{1}(kr)&\,e^{-\frac{r^2}{2\kappa}} j_{1}(k'r)\nonumber \\
&=&\frac{\sqrt{2\pi}}{4}\frac{1}{k^2 k'^2 \kappa^{1/2}}&\Big[\left(-1+k k' \kappa\right)e^{-\frac{\kappa}{2}(k-k')^2}+\left(1+k k' \kappa\right)e^{-\frac{\kappa}{2}(k+k')^2}\Big]\\[0.4cm] 
I_{\kappa}\left(k,k',2,2;0\right)&=&\int_{0}^{\infty} \mbox{d}r\, r^2 j_{2}(kr)&\,e^{-\frac{r^2}{2\kappa}} j_{2}(k'r)\nonumber \\
&=& \frac{\sqrt{2\pi}}{4}\frac{1}{k^3 k'^3 \kappa^{3/2}} &\Big[\left(3-3 k k' \kappa +(k k' \kappa )^2 \right)e^{-\frac{\kappa}{2}(k-k')^2} -\left(3+3k k' \kappa +(k k' \kappa )^2 \right)e^{-\frac{\kappa}{2}(k+k')^2} \Big]\\[0.3cm] 
I_{\kappa}\left(k,k',3,3;0\right)&=&\int_{0}^{\infty} \mbox{d}r\, r^2 j_{3}(kr)&\,e^{-\frac{r^2}{2\kappa}} j_{3}(k'r)\nonumber \\
&=& \frac{\sqrt{2\pi}}{4}\frac{1}{k^4 k'^4 \kappa^{5/2}} &\Big[\left(-15+15 k k' \kappa -6(k k' \kappa )^2+(k k' \kappa )^3 \right)e^{-\frac{\kappa}{2}(k-k')^2} \nonumber \\
&&&+\left(15+15 k k' \kappa +6(k k' \kappa )^2+(k k' \kappa )^3 \right)e^{-\frac{\kappa}{2}(k+k')^2}\Big]\\[0.3cm] 
I_{\kappa}\left(k,k',4,4;0\right)&=&\int_{0}^{\infty} \mbox{d}r\, r^2 j_{4}(kr)&\,e^{-\frac{r^2}{2\kappa}} j_{4}(k'r)\nonumber \\
&=& \frac{\sqrt{2\pi}}{4}\frac{1}{k^5 k'^5 \kappa^{7/2}}&\Big[\left(105-105 k k' \kappa  +45(k k' \kappa )^2 -10 (k k' \kappa )^3+(k k' \kappa )^4 \right)e^{-\frac{\kappa}{2}(k-k')^2}\nonumber \\
&&&-\left(105+105 k k' \kappa +45(k k' \kappa )^2+10 (k k' \kappa )^3+(k k' \kappa )^4 \right)e^{-\frac{\kappa}{4}(k+k')^2}\Big].
\end{align}
\label{eq:I0}
\end{subequations}
\end{widetext}
The expressions for $I_{\kappa}\left(k,k',L,L;n_P\!=2\right)$ can be obtained from those for $I_{\kappa}\left(k,k',L,L;n_P\!=0\right)$ in Eqs.~\eqref{eq:I0} by the relation
\begin{align}
I_{\kappa}\left(k,k',L,L;n_P\!=2\right)&=\nonumber \\
&-2\frac{\partial}{\partial\left(\kappa^{-1}\right)}I_{\kappa}\left(k,k',L,L;n_P\!=0\right).
\end{align}

The tensor operators in Eq.~\eqref{eq:ucompotential} and \eqref{eq:reducompotential} also connect states with angular momentum $L$ and $L'=L\pm2$, so that we need the integrals $I_{\kappa}\left(k,k',L,L+2;2\right)$:
\begin{widetext}
\begin{subequations}
\begin{align}
I_{\kappa}\left(k,k',0,2;2\right)&=&\int_{0}^{\infty} \mbox{d}r\, r^2 j_{0}(kr)&\,r^2e^{-\frac{r^2}{2\kappa}} j_{2}(k'r)\nonumber \\
&=& \frac{\sqrt{2\pi}}{4}\frac{1}{kk'^3\kappa^{-1/2}}&\Big[(3+2k'^2 \kappa +k'^4\kappa ^2-kk'\kappa(3+2k'^2\kappa )+(kk'\kappa )^2)e^{-\frac{\kappa}{2}(k-k')^2} \nonumber \\
&&&-(3+2k'^2 \kappa +k'^4\kappa ^2+kk'\kappa(3+2k'^2\kappa )+(kk'\kappa )^2)e^{-\frac{\kappa}{2}(k+k')^2}\Big]\\
I_{\kappa}\left(k,k',1,3;2\right)&=&\int_{0}^{\infty} \mbox{d}r\, r^2 j_{1}(kr)&\,r^2e^{-\frac{r^2}{2\kappa}} j_{3}(k'r)\nonumber \\
&=&
\frac{2\sqrt{\pi}}{4}\frac{1}{k^2k'^4\kappa^{1/2}}&\Big[\big(-(15+6k'^2\kappa +k'^4\kappa ^2)+kk'\kappa (15+6k'^2\kappa + k'^4 \kappa ^2)\nonumber \\
&&&-(kk'\kappa )^2(6+2k'^2\kappa )+(kk'\kappa)^3\big)e^{-\frac{\kappa}{2}(k-k')^2} \nonumber \\
&&&+\big((15+6k'^2\kappa +k'^4\kappa ^2)+kk'\kappa (15+6k'^2\kappa + k'^4 \kappa ^2)\nonumber \\
&&&+(kk'\kappa )^2(6+2k'^2\kappa )+(kk'\kappa)^3\big)e^{-\frac{\kappa}{2}(k+k')^2}\Big] \\
I_{\kappa}\left(k,k',2,4;2\right)&=&\int_{0}^{\infty} \mbox{d}r\, r^2 j_{2}(kr)&\,r^2e^{-\frac{r^2}{2\kappa}} j_{4}(k'r)\nonumber \\
&=&\frac{\sqrt{2\pi}}{4}\frac{1}{k^3k'^5\kappa^{3/2}}&\Big[\big((105+30k'^2\kappa+3k'^4\kappa ^2)-kk'\kappa(105+30k'^2\kappa+3k'^4\kappa ^2)\nonumber \\
&&&+(kk'\kappa)^2(45+8k'^2\kappa+k'^4\kappa ^2)-(kk'\kappa)^3(5+k'^2\kappa)\nonumber \\ 
&&&+(kk'\kappa)^4\big)e^{-\frac{\kappa}{2}(k-k')^2}\nonumber \\
&&&-\big((105+30k'^2\kappa+3k'^4\kappa ^2)+kk'\kappa(105+30k'^2\kappa+3k'^4\kappa ^2)\nonumber \\
&&&+(kk'\kappa)^2(45+8k'^2\kappa+k'^4\kappa ^2)+(kk'\kappa)^3(5+k'^2\kappa)\nonumber \\
&&&+(kk'\kappa)^4 \big)e^{-\frac{\kappa}{2}(k+k')^2}\Big]\\
I_{\kappa}\left(k,k',3,5;2\right)&=&\int_{0}^{\infty} \mbox{d}r\, r^2 j_{3}(kr)&\,r^2e^{-\frac{r^2}{2\kappa}} j_{5}(k'r)\nonumber \\
&=&\frac{\sqrt{2\pi}}{4}\frac{1}{k^4k'^6\kappa^{5/2}}&\Big[\big(-(945+210k'^2\kappa +15k'^4\kappa ^2)+kk'\kappa (945+210k'^2\kappa +15k'^4\kappa ^2) \nonumber \\
&&&-(kk'\kappa )^2(420+90k'^2\kappa +6k'^4\kappa ^2)+(kk'\kappa ^3)(105+20k'^2\kappa +k'^4\kappa^2)\nonumber \\
&&&-(kk'\kappa )^4(15+2k'^2 \kappa )+(kk'\kappa)^5)e^{-\frac{\kappa}{2}(k-k')^2}\nonumber \\
&&&+\big((945+210k'^2\kappa +15k'^4\kappa ^2)+kk'\kappa (945+210k'^\kappa +15k'^4\kappa ^2) \nonumber \\
&&&+(kk'\kappa )^2(420+90k'^2\kappa +6k'^4\kappa ^2)+(kk'\kappa ^3)(105+20k'^2\kappa +k'^4\kappa^2)\nonumber \\
&&&+(kk'\kappa )^4(15+2k'^2 \kappa )+(kk'\kappa)^5\big)e^{-\frac{\kappa}{4}(k+k')^2}\Big].
\end{align}
\label{eq:I2}
\end{subequations}
\end{widetext}
The integral
\begin{align}
&I'_{\kappa}\left(k,k',L,L'\right):=\nonumber \\
&\int_0^{\infty}drr^2 \left(\frac{\partial}{\partial r}rj_L(kr)\right)r^2\mbox{exp}\left\{-\frac{r^2}{2\kappa_{\mu}}\right\}j_{L'}(k'r)  \nonumber \\
&-\int_0^{\infty}drr^2 j_L(kr)r^2\mbox{exp}\left\{-\frac{r^2}{2\kappa_{\mu}}\right\}\left(\frac{\partial}{\partial r}rj_{L'}(k'r)\right), 
\end{align}
which is required for the matrix elements of the momentum dependent tensor term Eq.~\eqref{eq:ansatzmemom2} can be obtained from the expressions on the r.h.s. of Eqs.~\eqref{eq:I2} by calculating the following derivatives:
\begin{align}
I'_{\kappa}(k,k',L &,L+2)=\nonumber \\
&\left(k\frac{\partial}{\partial k}-k'\frac{\partial}{\partial k'}\right)I_{\kappa}\left(k,k',L,L+2;2\right).
\end{align}

\subsection{Operator matrix elements\label{sec:opme}}
In this section we list the partial wave matrix elements of the operators in the UCOM potential Eq.~(\ref{eq:ucompotential}). They are required to evaluate Eq.~(\ref{eq:ansatzme2}). For the quadratic angular momemtum operator and the spin-orbit operator one finds
\begin{eqnarray}
\bra{(LS)J}\vec{\op{L}}^{\,2}\ket{(L'S)J}&=&L(L+1))\delta_{LL'}\label{eq:opme1}\\
\bra{(LS)J}\vec{\op{L}}\cdot\vec{\op{S}}\ket{(L'S)J}&=&\frac{1}{2}\Big(J(J+1)-L(L+1)\nonumber \\
&&\quad-S(S+1)\Big)\delta_{LL'}.\label{eq:opme2}
\end{eqnarray}

The tensor operator $S_{12}(\vec{\op{L}},\vec{\op{L}})$ can be rewritten by means of Eq.~(\ref{eq:ls2}) as
\begin{eqnarray}
 S_{12}(\vec{\op{L}},\vec{\op{L}})=6(\vec{\op{L}}\cdot\vec{\op{S}})^2+3(\vec{\op{L}}\cdot\vec{\op{S}})-4\vec{\op{L}}^{\,2}\op{\Pi}_{S=1},\label{eq:tll}
\end{eqnarray}
and its matrix elements can be evaluated using Eqs.~(\ref{eq:opme1}) and (\ref{eq:opme2}).

The tensor operator $\op{S}_{12}$ connects not only states with equal $L$ but also $L$ and $L\pm2$. Its matrix elements are given in Tab.~\ref{tab:ten}.
\begin{table}[h]
\centering
\begin{ruledtabular}
\begin{tabular}{c|ccc}
$\bra{(L1)J}\op{S}_{12}\ket{(L'1)J}$& $L'=J-1$&$L'=J$&$L'=J+1$ \\ \hline 
$L=J-1$&$-\frac{2(J-1)}{2J+1}$&$0$&$\frac{6\sqrt{J(J+1)}}{2J+1}$\\ 
$L=J$&$0$&$2$&$0$\\ 
$L=J+1$&$\frac{6\sqrt{J(J+1)}}{2J+1}$&$0$&$-\frac{2(J+2)}{2J+1}$
\end{tabular}
\end{ruledtabular}
\caption{Matrix elements of the tensor operator $\op{S}_{12}$. For $S=0$ the matrix elements are zero.\label{tab:ten}}
\end{table} 

The tensor operators $\bar{S}_{12}(\vec{\op{p}}_{\Omega},\vec{\op{p}}_{\Omega})$ and $S_{12}(\vec{\op{r}},\vec{\op{p}_{\Omega}})$ connect only states with $L$ and $L'=L\pm2$. Its matrix elements are given by:
\begin{multline}
\bra{(J\!-\!1\,1)J}\bar{S}_{12}(\vec{\op{p}}_{\Omega},\vec{\op{p}}_{\Omega})\ket{(J\!+\!1\,1)J}=\\
-(3+6J)\sqrt{J(J+1)}
\end{multline}
\begin{multline}
\bra{(J\!-\!1\,1)J}S_{12}(\vec{\op{r}},\vec{\op{p}_{\Omega}})\ket{(J\!+\!1\,1)J}=\\
-3i\sqrt{J(J+1)}.
\end{multline}
\subsection{Separation of central, spin-orbit and tensor components \label{sec:separate}}
The operator representation of the UCOM transformed Argonne potential is obtained by fitting an ansatz for the operator representation to the partial wave matrix elements of the considered potential. One can classify the operators occurring in the ansatz by their spin dependence. The central part of the interaction consists of terms with operators which do not depend on the spin operator: $\op{1}$, $\vec{\op{L}}^2$ and $\vec{\op{p}}^{\,2}$. The spin-orbit part contains operators with tensor rank one in spin space, like $\vec{\op{L}}\cdot\vec{\op{S}}$ and $\vec{\op{L}}^2(\vec{\op{L}}\cdot\vec{\op{S}})$. The operators of the tensor part, e.g. $\op{S}_{12}$, $S_{12}(\vec{\op{L}},\vec{\op{L}})$, $\bar{S}_{12}(\vec{\op{p}}_{\Omega},\vec{\op{p}}_{\Omega})$ and $S_{12}(\vec{\op{r}},\vec{\op{p}_{\Omega}})$, are of rank two.

In the partial wave matrix elements with $S=0$, only the central part contributes. The $S=1$ matrix elements connecting different orbital angular momenta contain only tensor contributions. In the other cases, all three components are present. In that case one can separate the central, spin-orbit and tensor components for a given angular momentum $L$ by calculating linear combinations of the partial wave matrix elements with $J=L-1$, $J=L$ and $J=L+1$ \cite{Weber}.

The central part of the potential $\op{V}$ can be isolated by the linear combination
\begin{eqnarray}
\sum_{J=L-1}^{L+1}\frac{(2J+1)}{\sum_{J'=L-1}^{L+1}(2J'+1)}\bra{k(L1)J;T}\op{V}\ket{k'(L1)J;T}=\nonumber\\ \bra{k(L1)J;T}\Big(V^C_{1T}(\op{r})+V^{L2}_{1T}(\op{r})\vec{\op{L}}^2\nonumber\\
+\frac{1}{2}\left[\vec{\op{p}}^{\,2}V_{1T}^{p2}(\op{r}) + V_{1T}^{p2}(\op{r})\vec{\op{p}}^{\,2}\right]\Big)\ket{k'(L1)J;T}.\nonumber \\\label{eq:isolation}
\end{eqnarray}
Using the parameters
\begin{eqnarray}
\alpha^L_J=\left\{
\begin{array}{*{2}{cl}}
-\frac{2L^2+L-1}{4 L^3+6L^2+2L},& J=L-1 \\
-\frac{2L+1}{4 L^3+6L^2+2L}, &J=L \\
\frac{2L^2+3L}{4 L^3+6L^2+2L}, &J=L+1\\
\end{array}
\right.
\end{eqnarray}
one obtains the spin-orbit contributions
\begin{eqnarray}
\sum_{J=L-1}^{L+1}\alpha^L_J\,\bra{k(L1)J;T}\op{V}\ket{k'(L1)J;T}=\nonumber\\
\bra{k(L1)J;T}V^{LS}_{1T}(\op{r})\nonumber+V^{L2LS}_{1T}(\op{r})\vec{\op{L}}^2+\cdots\ket{k'(L1)J;T}.\nonumber\\
\end{eqnarray}
For the tensor part, we can use for example
\begin{eqnarray}
\beta^{L}_{J}=\left\{
\begin{array}{*{2}{cl}}
\frac{L+1}{12L^3+18L^2+6L},\qquad J=L-1 \\
-\frac{2L+1}{12L^3+18L^2+6L},\qquad J=L \\
\frac{L}{12L^3+18L^2+6L},\qquad J=L+1\\
\end{array}
\right.
\end{eqnarray}
and we obtain
\begin{eqnarray}
\sum_{J=L-1}^{L+1}\beta^L_J\,\bra{k(L1)J;T}\op{V}\ket{k'(L1)J;T}= \nonumber \\
\bra{k(L1)J;T}\Big(-\frac{2}{(2L+3)(2L-1)}V^{T}_{1T}(\op{r})\nonumber \\
+V^{Tll}_{1T}(\op{r})+\cdots\Big)\ket{k'(L1)J;T}.
\end{eqnarray}
By using this technique, it is possible to fit individually the central, spin-orbit and tensor component in the ansatz of the operator representation to the linear combined matrix elements containing only the desired component of the interaction.

\subsection{Weights\label{Sec:Weight}}
The fitting method to derive the operator representation allows the partial waves included in the fit to be weighed differently. The weight factors are chosen such that the lowest angular momentum partial wave matrix elements are reproduced in an optimal way and the deviations in the partial waves with higher $L$ remain as small as possible. Tabs.~\ref{tab:weightreducom1} and \ref{tab:weightreducom2} show the weight factors that were used in the fits to obtain the operator representations. For $S=1$, central, spin-orbit and tensor part were fitted separately (see App.~\ref{sec:separate}), so that the weight factors for each of these fits are given separately as well.

\begin{table}[h]
\centering
\begin{ruledtabular}
\begin{tabular}{c|c|ccc}
pw's with & $S=0$, $T=1$ &  \multicolumn{3}{c}{$S=1$, $T=0$}  \\ 
$L$& central & \multicolumn{1}{c}{central} & \multicolumn{1}{c}{spin-orbit} & \multicolumn{1}{c}{tensor}\\ \hline
0&1&1&-&-\\
0-2&-&-&-&0.1\\
2&1&1&1&1\\
2-4&-&-&-&0.01\\
4&0.1&0.1&0.2&0.2
\end{tabular}
\end{ruledtabular}
\caption{\label{tab:weightreducom1} Weight factors for matrix elements with even angular momentum $L$ used in the fit for the operator representation of the reduced UCOM potential.}
\end{table}

\begin{table}[h]
\centering
\begin{ruledtabular}
\begin{tabular}{c|c|ccc}
pw's with & $S=0$, $T=0$ &  \multicolumn{3}{c}{$S=0$, $T=1$}  \\ 
$L$& central & \multicolumn{1}{c}{central} & \multicolumn{1}{c}{spin-orbit} & \multicolumn{1}{c}{tensor}\\ \hline
1&1&1&1&1\\
1-3&-&-&-&0.01\\
3&1&1&0.1&0.01\\ 
3-5&-&-&-&0.01
\end{tabular}
\end{ruledtabular}
\caption{\label{tab:weightreducom2} Weight factors for matrix elements with odd angular momentum $L$ used in the fit for the operator representation of the reduced UCOM potential.}
\end{table}

\section{Parameterization of the radial functions \label{sec:parameters}}
The local radial functions $\mathcal{V}^{P}_{ST}(r)$ of the reduced UCOM potential Eq.~(\ref{eq:reducompotential}) described in Sec.~\ref{sec:op} are parameterized by a sum of Gaussians
\begin{subequations}
\begin{eqnarray} 
\mathcal{V}^P_{ST}(r)=\sum_{\mu}\gamma^{P}_{ST,\mu}\mbox{exp}\Big\{-\frac{r^{\,2}}{2\kappa_{\mu}}\Big\},\label{eq:paraloc}
\end{eqnarray}
with $P\in\{C,\,L2,\,p2,\,LS,\,Tll\}$ and
\begin{eqnarray} 
\mathcal{V}^T_{1T}(r)&=&\sum_{\mu}\gamma^{T}_{1T,\mu}r^2\cdot\mbox{exp}\Big\{-\frac{r^{\,2}}{2\kappa_{\mu}}\Big\} \\
\mathcal{V}^{Trq}_{1T}(r)&=&\sum_{\mu}\gamma^{Trq}_{1T,\mu}r^3\cdot\mbox{exp}\Big\{-\frac{r^{\,2}}{2\kappa_{\mu}}\Big\}.\label{eq:paraten}
\end{eqnarray}
\label{eq:paralocall}
\end{subequations}
The parameters $\kappa$ are chosen by the relation
\begin{eqnarray*} 
\kappa_{\mu}=\kappa_1 \cdot b^{\mu-1},
\end{eqnarray*} 
with $\kappa_1=0.05\,\mbox{fm}^2$, $b=2$ and $\mu_{\rm{max}}=8$, which corresponds to a maximum width parameter $\kappa_{8}=6.4\,\mbox{fm}^2$. With these parameters, one is able to cover the whole range of the interaction, which is a few $\mbox{fm}$. The parameters $\gamma^{P}_{ST,\mu}$ obtained by the fit described in Sec.~\ref{sec:op} for the UCOM(SRG) transformed Argonne potential with a flow parameter of $0.04\,\mathrm{fm}^4$ are presented in Tab.~\ref{tab:st00} - \ref{tab:st11}. The results for the UCOM(SRG) transformed Argonne potential with $\alpha=0.2\,\mathrm{fm}^4$, which is more suitable for FMD and AMD calculations, are presented in Tabs.~\ref{tab:2000st00} - \ref{tab:2000st11}.

\begin{table*}[h]
\centering
\begin{ruledtabular}
\begin{tabular}{c .|*{3}{.}|*{3}{.}}
\multicolumn{1}{c}{\#} & \multicolumn{1}{c|}{$\kappa_{\mu}\,[\mbox{fm}^2]$} &\multicolumn{1}{c}{$\gamma^{C}_{00,\mu}\,[\mbox{MeV}]$} & \multicolumn{1}{c}{$\gamma^{L2}_{00,\mu}\,[\mbox{MeV}]$} & \multicolumn{1}{c|}{$\gamma^{p2}_{00,\mu}\,[\mbox{MeV}\,\mbox{fm}^2]$} & \multicolumn{1}{c}{$\gamma^{C}_{01,\mu}\,[\mbox{MeV}]$} & \multicolumn{1}{c}{$\gamma^{L2}_{01,\mu}\,[\mbox{MeV}]$} & \multicolumn{1}{c}{$\gamma^{p2}_{01,\mu}\,[\mbox{MeV}\,\mbox{fm}^2]$} \\ \hline
1	&	0.05	&	3.3309	&	0.3120	&	-0.0317	&	12.7489	&	-0.9319	&	-0.1567	\\ 
2	&	0.1	&	1.4133	&	0.0115	&	-0.1234	&	-4.0758	&	-1.1318	&	-0.2739	\\ 
3	&	0.2	&	0.3816	&	-0.0595	&	-0.0510	&	2.5087	&	-0.5674	&	-0.0228	\\ 
4	&	0.4	&	0.1905	&	-0.0984	&	-0.0579	&	-3.5183	&	-0.3815	&	-0.0742	\\ 
5	&	0.8	&	-0.1135	&	-0.0570	&	0.0757	&	0.8700	&	0.0131	&	0.5872	\\ 
6	&	1.6	&	0.1222	&	-0.0038	&	0.0669	&	-0.0852	&	0.0258	&	-0.1897	\\ 
7	&	3.2	&	0.0013	&	0.0006	&	-0.0139	&	-0.0113	&	-0.0014	&	-0.0150	\\
8	&	6.4	&	0.0094	&	-0.0002	&	0.0031	&	-0.0041	&	-0.0002	&	0.0080	
\end{tabular}
\end{ruledtabular}
\caption{\label{tab:st00} The parameters $\gamma^{P}_{ST,\mu}$ of the reduced UCOM(0.04) fit ($\alpha=0.04\,\mathrm{fm}^4$) for both $S=0$ channels.}
\vspace{2cm}
\centering
\begin{ruledtabular}
\begin{tabular}{c .|*{7}{c}}
\multicolumn{1}{c}{\#} & \multicolumn{1}{c|}{$\kappa_{\mu}\,[\mbox{fm}^2]$} & \multicolumn{1}{c}{$\gamma^{C}_{10,\mu}\,[\mbox{MeV}]$} & \multicolumn{1}{c}{$\gamma^{L2}_{10,\mu}\,[\mbox{MeV}]$} & \multicolumn{1}{c}{$\gamma^{p2}_{10,\mu}\,[\mbox{MeV}\,\mbox{fm}^2]$} & \multicolumn{1}{c}{$\gamma^{LS}_{10,\mu}\,[\mbox{MeV}]$} & \multicolumn{1}{c}{$\gamma^{T}_{10,\mu}\,[\mbox{MeV}\,\mbox{fm}^{-2}]$} & \multicolumn{1}{c}{$\gamma^{Tll}_{10,\mu}\,[\mbox{MeV}]$} & \multicolumn{1}{c}{$\gamma^{Trp}_{10,\mu}\,[\mbox{MeV}\,\mbox{fm}^{-2}]$} \\ \hline
1	&	0.05	&	15.0990	&	-0.0805	&	-0.2054	&	0	&	5.6641	&	0	&	-1.1195	\\ 
2	&	0.1	&	-9.9144	&	-1.3637	&	-0.1667	&	0.8178	&	-2.4454	&	0.0290	&	-0.0793	\\ 
3	&	0.2	&	7.6191	&	-0.2134	&	0.0028	&	0.1128	&	0.6413	&	-0.1042	&	0.1874	\\ 
4	&	0.4	&	-5.6983	&	-0.2714	&	-0.3437	&	0.4323	&	-0.3009	&	0.036	&	0.0059	\\ 
5	&	0.8	&	1.1003	&	0.0311	&	0.7655	&	-0.2032	&	-0.0118	&	0.0174	&	0.0326	\\ 
6	&	1.6	&	-0.0645	&	0.0281	&	-0.1433	&	0.0560	&	-0.0192	&	-0.0014	&	-0.0017	 \\ 
7	&	3.2	&	0.0016	&	0.0026	&	-0.0863	&	-0.0071	&	0.0003	&	-0.0003	&	-0.0001	 \\ 
8	&	6.4	&	-0.0091	&	-0.0010	&	0.0239	&	0.0008	&	-0.0002	&	0.0001	&	0.0000	
\end{tabular}
\end{ruledtabular}
\caption{\label{tab:st10} The parameters $\gamma^{P}_{10,\mu}$ of the reduced UCOM(0.04) fit ($\alpha=0.04\,\mathrm{fm}^4$) for $S=1$ and $T=0$. The parameters $\gamma^{LS}_{10,1}$ and $\gamma^{Tll}_{10,1}$ are set to zero to improve the stability of the fit.}
\vspace{2cm}
\centering
\begin{ruledtabular}
\begin{tabular}{c .|*{7}{.}}
\multicolumn{1}{c}{\#} & \multicolumn{1}{c|}{$\kappa_{\mu}\,[\mbox{fm}^2]$} & \multicolumn{1}{c}{$\gamma^{C}_{11,\mu}\,[\mbox{MeV}]$} & \multicolumn{1}{c}{$\gamma^{L2}_{11,\mu}\,[\mbox{MeV}]$} & \multicolumn{1}{c}{$\gamma^{p2}_{11,\mu}\,[\mbox{MeV}\,\mbox{fm}^2]$} & \multicolumn{1}{c}{$\gamma^{LS}_{11,\mu}\,[\mbox{MeV}]$} & \multicolumn{1}{c}{$\gamma^{T}_{11,\mu}\,[\mbox{MeV}\,\mbox{fm}^{-2}]$} & \multicolumn{1}{c}{$\gamma^{Tll}_{11,\mu}\,[\mbox{MeV}]$} & \multicolumn{1}{c}{$\gamma^{Trp}_{11,\mu}\,[\mbox{MeV}\,\mbox{fm}^{-2}]$} \\ \hline
1	&	0.05	&	6.6519	&	0.0504	&	-0.0539	&	0	&	0	&	0	&	-1.1195	\\ 
2	&	0.1	&	1.8263	&	0.0122	&	-0.1984	&	-0.4740	&	4.2475	&	0	&	-0.0793	\\ 
3	&	0.2	&	-2.9392	&	0.9389	&	0.0183	&	-2.5248	&	-0.2895	&	0	&	0.1874	\\ 
4	&	0.4	&	0.2188	&	-0.1731	&	0.0070	&	-0.0228	&	0.1699	&	0	&	0.0059	\\ 
5	&	0.8	&	-0.0579	&	-0.0403	&	0.0889	&	-0.0333	&	-0.0026	&	0	&	0.0326	\\ 
6	&	1.6	&	0.0362	&	0.0040	&	0.0248	&	0.0090	&	0.0067	&	0	&	-0.0017	\\ 
7	&	3.2	&	-0.0036	&	0.0003	&	-0.0148	&	-0.0025	&	-0.0001	&	0	&	-0.0001	\\ 
8	&	6.4	&	0.0009	&	-0.0001	&	0.0026	&	0.0003	&	0.0001	&	0	&	0.0000	
\end{tabular}
\end{ruledtabular}
\caption{\label{tab:st11} The parameters $\gamma^{P}_{11,\mu}$ of the reduced UCOM(0.04) fit ($\alpha=0.04\,\mathrm{fm}^4$) for $S=1$ and $T=1$. The parameters $\gamma^{LS}_{11,1}$, $\gamma^{T}_{11,1}$ and $\gamma^{Tll}_{11,\mu}$ are set to zero to improve the stability of the fit.}
\end{table*}

\begin{table*}[h!]
\centering
\begin{ruledtabular}
\begin{tabular}{c .|*{3}{.}|*{3}{.}}
\multicolumn{1}{c}{\#} & \multicolumn{1}{c|}{$\kappa_{\mu}\,[\mbox{fm}^2]$} &\multicolumn{1}{c}{$\gamma^{C}_{00,\mu}\,[\mbox{MeV}]$} & \multicolumn{1}{c}{$\gamma^{L2}_{00,\mu}\,[\mbox{MeV}]$} & \multicolumn{1}{c|}{$\gamma^{p2}_{00,\mu}\,[\mbox{MeV}\,\mbox{fm}^2]$} & \multicolumn{1}{c}{$\gamma^{C}_{01,\mu}\,[\mbox{MeV}]$} & \multicolumn{1}{c}{$\gamma^{L2}_{01,\mu}\,[\mbox{MeV}]$} & \multicolumn{1}{c}{$\gamma^{p2}_{01,\mu}\,[\mbox{MeV}\,\mbox{fm}^2]$} \\ \hline
1	&	0.05	&	3.0992	&	0.1797	&	-0.0372	&	6.1774	&	-1.3066	&	-0.0511	\\ 
2	&	0.1	&	1.5233	&	0.0631	&	-0.0998	&	6.2590	&	-0.5749	&	-0.3748	\\ 
3	&	0.2	&	0.1075	&	-0.0952	&	-0.0578	&	-4.9098	&	-1.0211	&	-0.1570	\\ 
4	&	0.4	&	0.2470	&	-0.0802	&	-0.0233	&	0.5672	&	-0.1991	&	0.2560	\\ 
5	&	0.8	&	-0.0627	&	-0.0556	&	0.0206	&	-0.9899	&	-0.1095	&	-0.0039	\\ 
6	&	1.6	&	0.0601	&	-0.0138	&	0.0279	&	0.3929	&	0.0453	&	0.5148	\\ 
7	&	3.2	&	0.0089	&	-0.0021	&	0.0314	&	-0.0433	&	0.0101	&	-0.3464	\\ 
8	&	6.4	&	0.0112	&	0.0000	&	0.0023	&	-0.0107	&	-0.0016	&	0.0518
\end{tabular}
\end{ruledtabular}
\caption{\label{tab:2000st00} The parameters $\gamma^{P}_{ST,\mu}$ of the reduced UCOM(0.20) potential ($\alpha=0.2\,\mathrm{fm}^4$) for both $S=0$ channels.}
\vspace{2cm}
\centering
\begin{ruledtabular}
\begin{tabular}{c .|*{7}{.}}
\multicolumn{1}{c}{\#} & \multicolumn{1}{c|}{$\kappa_{\mu}\,[\mbox{fm}^2]$} & \multicolumn{1}{c}{$\gamma^{C}_{10,\mu}\,[\mbox{MeV}]$} & \multicolumn{1}{c}{$\gamma^{L2}_{10,\mu}\,[\mbox{MeV}]$} & \multicolumn{1}{c}{$\gamma^{p2}_{10,\mu}\,[\mbox{MeV}\,\mbox{fm}^2]$} & \multicolumn{1}{c}{$\gamma^{LS}_{10,\mu}\,[\mbox{MeV}]$} & \multicolumn{1}{c}{$\gamma^{T}_{10,\mu}\,[\mbox{MeV}\,\mbox{fm}^{-2}]$} & \multicolumn{1}{c}{$\gamma^{Tll}_{10,\mu}\,[\mbox{MeV}]$} & \multicolumn{1}{c}{$\gamma^{Trp}_{10,\mu}\,[\mbox{MeV}\,\mbox{fm}^{-2}]$} \\ \hline
1	&	0.05	&	-2.3036	&	-1.3549	&	0.0800	&	0	&	1.9680	&	0	&	-3.2469	\\
2	&	0.1	&	17.3967	&	0.3762	&	-0.4563	&	0.9670	&	-0.2879	&	0.0676	&	0.6973	\\
3	&	0.2	&	-13.0260	&	-1.4853	&	-0.2465	&	0.0135	&	-0.1399	&	-0.1401	&	-0.1998	\\
4	&	0.4	&	5.7847	&	0.1845	&	0.4814	&	0.2761	&	-0.0116	&	0.0269	&	0.0983	\\
5	&	0.8	&	-3.6956	&	-0.1255	&	-0.5939	&	0.0966	&	-0.0411	&	0.0196	&	-0.0143	\\
6	&	1.6	&	1.0458	&	0.0768	&	1.2315	&	-0.0035	&	0.0011	&	0.0157	&	0.0106	\\
7	&	3.2	&	-0.1101	&	0.0190	&	-0.6762	&	0.0073	&	-0.0034	&	-0.0031	&	-0.0008	\\
8	&	6.4	&	-0.0098	&	-0.0025	&	0.0980	&	0.0009	&	0.0001	&	0.0003	&	0.0000
\end{tabular}
\end{ruledtabular}
\caption{\label{tab:2000st10} The parameters $\gamma^{P}_{10,\mu}$ of the reduced UCOM(0.20) fit ($\alpha=0.2\,\mathrm{fm}^4$) for $S=1$ and $T=0$. The parameters $\gamma^{LS}_{10,1}$ and $\gamma^{Tll}_{10,1}$ are set to zero to improve the stability of the fit.}
\vspace{2cm}
\centering
\begin{ruledtabular}
\begin{tabular}{c .|*{7}{.}}
\multicolumn{1}{c}{\#} & \multicolumn{1}{c|}{$\kappa_{\mu}\,[\mbox{fm}^2]$} & \multicolumn{1}{c}{$\gamma^{C}_{11,\mu}\,[\mbox{MeV}]$} & \multicolumn{1}{c}{$\gamma^{L2}_{11,\mu}\,[\mbox{MeV}]$} & \multicolumn{1}{c}{$\gamma^{p2}_{11,\mu}\,[\mbox{MeV}\,\mbox{fm}^2]$} & \multicolumn{1}{c}{$\gamma^{LS}_{11,\mu}\,[\mbox{MeV}]$} & \multicolumn{1}{c}{$\gamma^{T}_{11,\mu}\,[\mbox{MeV}\,\mbox{fm}^{-2}]$} & \multicolumn{1}{c}{$\gamma^{Tll}_{11,\mu}\,[\mbox{MeV}]$} & \multicolumn{1}{c}{$\gamma^{Trp}_{11,\mu}\,[\mbox{MeV}\,\mbox{fm}^{-2}]$} \\ \hline
1	&	0.05	&	6.1761	&	-0.1569	&	-0.0486	&	0	&	0	&	0	&	-3.2469	\\ 
2	&	0.1	&	2.1816	&	0.2639	&	-0.1902	&	-0.5655	&	4.2295	&	0	&	0.6973	\\
3	&	0.2	&	-3.3880	&	0.8520	&	0.0056	&	-2.4264	&	-0.3211	&	0	&	-0.1998	\\
4	&	0.4	&	0.4997	&	-0.1526	&	0.0375	&	-0.0015	&	0.1726	&	0	&	0.0983	\\
5	&	0.8	&	-0.1342	&	-0.0472	&	0.0342	&	-0.0368	&	-0.0033	&	0	&	-0.0143	\\ 
6	&	1.6	&	0.0320	&	0.0010	&	0.0482	&	0.0084	&	0.0068	&	0	&	0.0106	\\
7	&	3.2	&	0.0011	&	0.0004	&	-0.0102	&	-0.0025	&	-0.0001	&	0	&	-0.0008	\\
8	&	6.4	&	0.0006	&	0.0000	&	0.0006	&	0.0003	&	0.0001	&	0	&	0.0000
\end{tabular}
\end{ruledtabular}
\caption{\label{tab:2000st11} The parameters $\gamma^{P}_{11,\mu}$ of the reduced UCOM(0.20) fit ($\alpha=0.2\,\mathrm{fm}^4$) for $S=1$ and $T=1$. The parameters $\gamma^{LS}_{11,1}$, $\gamma^{T}_{11,1}$ and $\gamma^{Tll}_{11,\mu}$ are set to zero to improve the stability of the fit.}
\end{table*}
\clearpage

\section{FMD matrix elements \label{sec:fmdme}}
In this section, we present the matrix elements of the reduced UCOM potential Eq.~(\ref{eq:reducompotential}), using the FMD single-particle states
\begin{eqnarray}
\ket{q_k}=\ket{a_k\,\vec{b}_k}\otimes\ket{\chi_k}
\end{eqnarray}
with 
\begin{eqnarray}
\braket{\vec{r}}{a_k\,\vec{b}_k}={\rm exp}\Big\{-\frac{1}{2a_k}(\vec{r}-\vec{b}_k)^2\Big\}
\end{eqnarray}
and the two-component spinor $\chi_k$. We use the abbreviations:
\begin{eqnarray*}
\lambdakl &=&\frac{1}{\ack+\al}\\
\alphakl &=&\frac{\ack\al}{\ack+\al}\\
\pikl &=&i \frac{\bck-\bl}{\ack+\al}\\
\rhokl &=& \frac{\al\bck+\ack\bl}{\ack+\al}\\
R_{kl}&=&\braket{\ak\bk}{\al\bl}=(2\pi\alphakl)^{3/2}{\rm exp}\Big\{\frac{\pikl^2}{2\lambdakl}\Big\}
\end{eqnarray*}
and
\begin{eqnarray*}
\lambdaklmn &=&\lambdakm + \lambdaln\\
\alphaklmn &=&\alphakm+\alphaln\\
\piklmn &=& \frac{1}{2}(\pikm-\piln)\\
\rhoklmn &=& \rhokm-\rholn\\
\betaklmn&=&i[(\ack-\am)\lambdakm+(\acl-\an)\lambdaln]\\
\thetaklmn&=&(\ack\lambdakm+\acl\lambdaln)(\am\lambdakm+\an\lambdaln).
\end{eqnarray*}
Since the radial functions are parameterized by a sum of Gaussians (Eq.~\eqref{eq:parameterization}), the two-body matrix elements of the reduced UCOM potential in the FMD basis are given by 
\begin{align}
&\bra{q_k,q_l}\op{V}^{(\rm{red.})}_{\rm{UCOM}}\ket{q_m,q_n}=\sum_p\sum_{ST}\sum_{\mu}\frac{1}{2}\gamma^{P}_{ST,\,\mu}\cdot\nonumber \\
&\bra{q_k,q_l}\!\!\left(\!\op{\mathcal{O}}_P\,\op{r}^{n^P}\!G_{\mu}(\op{r})+\op{\mathcal{O}}_P\,\op{r}^{n^P}\!G_{\mu}(\op{r})\op{\mathcal{O}}_P\!\right)\!\op{\Pi}_{ST}\ket{q_m,q_n},\nonumber \\
\end{align}
so that for all operators $\op{\mathcal{O}}_P$ of the reduced UCOM potential the expressions for the matrix elements
\begin{eqnarray}
\bra{q_k,q_l}\op{\mathcal{O}}_P\,\op{r}^{n^P}\!G_{\mu}(\op{r})+\op{\mathcal{O}}_P\,\op{r}^{n^P}\!G_{\mu}(\op{r})\op{\mathcal{O}}_P\ket{q_m,q_n}\label{eq:meFMD}
\end{eqnarray}
for a standard set of Gaussians $G_{\mu}(r)=\mbox{exp}\left\{-\frac{r^2}{2\kappa_{\mu}}\right\}$ with the width $\kappa_{\mu}$ have to be calculated. The FMD basis matrix element of a Gaussian is given by  \cite{Neffdipl}:
\begin{align}
G_{klmn}^{\,\mu}&=\bra{\ak\bk,\al\bl}G_{\mu}(\op{r})\ket{\am\vec{b}_m,\an\bn}=\nonumber\\
= R&_{km} R_{ln} \left(\frac{\kappa_{\mu}}{\alphaklmn+\kappa_{\mu}}\right)^{3/2}\mathrm{exp}\left\{\frac{-\rhoklmn^{\,2}}{2(\alphaklmn+\kappa_{\mu})}\right\}.
\end{align}
With the spin matrix element
\begin{align}
\vec{\sigma}_{kl}=\bra{\chi_k}\vec{\op{\sigma}}\ket{\chi_l}
\end{align}
and the definitions
\begin{align}
\vec{S}_{klmn} = \bra{\chi_k,\chi_l}\frac{1}{2}(&\vec{\op{\sigma}}(1)+\vec{\op{\sigma}}(2))\ket{\chi_m,\chi_n}\\
S_{12}(\vec{v}_{klmn},\vec{w}_{klmn}) =& \frac{3}{2}\Big\{(\vec{\sigma}_{km}\cdot\vec{v}_{klmn})(\vec{\sigma}_{ln}\cdot\vec{w}_{klmn})\nonumber \\
 &+(\vec{\sigma}_{km}\cdot\vec{w}_{klmn})(\vec{\sigma}_{ln}\cdot\vec{v}_{klmn})\Big\}\nonumber \\  
 &-(\vec{\sigma}_{km}\cdot\vec{\sigma}_{ln})(\vec{v}_{klmn}\cdot\vec{w}_{klmn}),
\end{align}
one finds for the parameterization Eq.~(\ref{eq:parameterization}): 
\begin{widetext}
\begin{align}
\bra{a_k\vec{b}_k,a_l\vec{b}_l}G_{\mu}(\op{r})\vec{\op{L}}^{\,2}\ket{a_m\vec{b}_m,a_n\vec{b}_n}=&\frac{\kappa_{\mu}}{\alphaklmn+\kappa_{\mu}}\Bigg\{\frac{\kappa_{\mu}(\rhoklmn\times\piklmn)^2}{\alphaklmn+\kappa_{\mu}}+\frac{1}{2}\frac{\lambdaklmn\kappa_{\mu}+\thetaklmn}{\alphaklmn+\kappa_{\mu}}\rhoklmn^{\,2}  \nonumber \\
+2\alpha\piklmn^{\,2}-&\betaklmn\rhoklmn\cdot\piklmn  -\frac{3}{2}(\thetaklmn-\alphaklmn\lambdaklmn)\Bigg\}G_{klmn}^{\,\mu} \\
\bra{a_k\vec{b}_k,a_l\vec{b}_l}\frac{1}{2}\Big[\vec{\op{p}}^{\,2}\,G_{\mu}(\op{r})+G_{\mu}(\op{r})\vec{\op{p}}^{\,2}\Big]\ket{a_m\vec{b}_m,a_n\vec{b}_n}=&\Bigg\{\piklmn^2-\frac{1}{2}\frac{\betaklmn}{\alphaklmn+\kappa_{\mu}}\rhoklmn\cdot\piklmn\nonumber\\ 
+\frac{1}{4}&\frac{\thetaklmn}{(\alphaklmn+\kappa_{\mu})^2}\rhoklmn^{\,2}+\frac{3}{4}(\lambdaklmn-\frac{\thetaklmn}{\alphaklmn+\kappa_{\mu}})\Bigg\}\,G_{klmn}^{\,\mu}  \\
\bra{a_k\vec{b}_k\chi_k,a_l\vec{b}_l\chi_l}G_{\mu}(\op{r})(\vec{\op{L}}\cdot\vec{\op{S}})\ket{a_m\vec{b}_m\chi_m,a_n\vec{b}_n\chi_n}=&\frac{\kappa_{\mu}(\rhoklmn\times\piklmn)\cdot\vec{S}_{klmn}}{\alphaklmn+\kappa_{\mu}}\,G_{klmn}^{\,\mu}
\end{align}
\end{widetext}
\begin{widetext}
\begin{align}
\bra{a_k\vec{b}_k\chi_k,a_l\vec{b}_l\chi_l}\op{r}^{\,2}G_{\mu}(\op{r})\op{S}_{12}\ket{a_m\vec{b}_m\chi_m,a_n\vec{b}_n\chi_n}=&\left(\frac{\kappa_{\mu}}{\alphaklmn+\kappa_{\mu}}\right)^2\,S_{12}(\rhoklmn,\rhoklmn)\,G_{klmn}^{\,\mu} 
\end{align}
\begin{align}
\bra{a_k\vec{b}_k\chi_k,a_l\vec{b}_l\chi_l}G_{\mu}(\op{r})S_{12}(\vec{\op{L}},\vec{\op{L}})\ket{a_m\vec{b}_m\chi_m,a_n\vec{b}_n\chi_n}=\frac{\kappa_{\mu}}{\alphaklmn+\kappa_{\mu}}\Bigg\{\frac{\kappa_{\mu}S_{12}(\rhoklmn\times\piklmn,\rhoklmn\times\piklmn)}{\alphaklmn+\kappa_{\mu}}
\nonumber \\
-\frac{1}{4}\frac{\lambdaklmn\kappa_{\mu}+\thetaklmn}{\alphaklmn+\kappa_{\mu}}\,S_{12}(\rhoklmn,\rhoklmn)-\alphaklmn\,S_{12}(\piklmn,\piklmn) +\frac{1}{2}\betaklmn\,S_{12}(\rhoklmn,\piklmn)\Bigg\}\,G_{klmn}^{\,\mu},
\end{align}

\begin{align}
\bra{a_k\vec{b}_k\chi_k,a_l\vec{b}_l\chi_l}\frac{1}{2}\Big[\op{p}_r\op{r}^3G_{\mu}(\op{r})+\op{r}^3G_{\mu}(\op{r})\op{p}_r\Big]S_{12}(\vec{\op{r}},\vec{\op{p}_{\Omega}})\ket{a_m\vec{b}_m\chi_m,a_n\vec{b}_n\chi_n}=\frac{1}{2}\left(\frac{\kappa_{\mu}}{\alphaklmn+\kappa_{\mu}}\right)^2\cdot\nonumber \\
\Bigg\{S_{12}(\rhoklmn,\rhoklmn)\bigg(\frac{1}{2}\frac{\kappa_{\mu}^2\betaklmn\,\rhoklmn^{\,2}\rhoklmn\cdot\piklmn}{(\alphaklmn+\kappa_\mu)^3}-2\left(\frac{\kappa_{\mu}\,\rhoklmn\cdot\piklmn}{\alphaklmn+\kappa_{\mu}}\right)^2+\frac{3}{8}\frac{\kappa_{\mu}\betaklmn^2\,\rhoklmn^{\,2}}{(\alphaklmn+\kappa_{\mu})^2}\nonumber \\
-\frac{1}{2}\left(1+9\frac{\kappa_{\mu}}{\alphaklmn+\kappa_{\mu}}\right)\frac{\kappa_{\mu}\betaklmn\,\rhoklmn\cdot\piklmn}{\alphaklmn+\kappa_{\mu}}-2\frac{\alphaklmn\kappa_{\mu}\,\piklmn^2}{\alphaklmn+\kappa_{\mu}}+\frac{3}{2}\thetaklmn-\frac{21}{8}\frac{\kappa_{\mu}\betaklmn^2}{\alphaklmn+\kappa_{\mu}}-3\bigg)\nonumber\\
+S_{12}(\rhoklmn,\piklmn)\bigg(-\frac{1}{2}\frac{\kappa_{\mu}^2\betaklmn\left(\rhoklmn^{\,2}\right)^2}{(\alphaklmn+\kappa_{\mu})^3}+2\frac{\kappa_{\mu}^2\rhoklmn^{\,2}(\rhoklmn\cdot\piklmn)}{(\alphaklmn+\kappa_{\mu})^2}\nonumber \\
 -\left(\frac{5}{2}-\frac{6\kappa_{\mu}}{\alphaklmn+\kappa_{\mu}}\right)\frac{\kappa_{\mu}\betaklmn\rhoklmn^{\,2}}{\alphaklmn+\kappa_{\mu}}+6\frac{\alphaklmn\kappa_{\mu}\,\rhoklmn\cdot\piklmn}{\alphaklmn+\kappa_{\mu}}-\frac{3}{2}\left(2-7\frac{\kappa_{\mu}}{\alphaklmn+\kappa_{\mu}}\right)\alphaklmn\betaklmn\bigg)\nonumber\\
 +2\alphaklmn S_{12}(\piklmn,\piklmn)\bigg(\frac{\kappa_{\mu}\rhoklmn^{\,2}}{\alphaklmn+\kappa_{\mu}}+3\alphaklmn\bigg)\Bigg\}\,G_{klmn}^{\,\mu}.
\end{align}
\end{widetext}
For completeness we also present the matrix elements for the other operators of the UCOM potential Eq.~(\ref{eq:ucompotential}): 
\begin{widetext} 
\begin{align}
&\bra{a_k\vec{b}_k\chi_k,a_l\vec{b}_l\chi_l}G_{\mu}(\op{r})\vec{\op{L}}^{\,2}(\vec{\op{L}}\cdot\vec{\op{S}})\ket{a_m\vec{b}_m\chi_m,a_n\vec{b}_n\chi_n}=\frac{\kappa_{\mu}(\rhoklmn\times\piklmn)\cdot\vec{S}_{klmn}}{\alphaklmn+\kappa_{\mu}}\Bigg\{\frac{\kappa_{\mu}^2(\rhoklmn\times\piklmn)^2}{(\alphaklmn+\kappa_{\mu})^2}\nonumber \\
&+\frac{\kappa_{\mu}\left(\lambdaklmn\kappa_{\mu}+\thetaklmn\right)}{(\alphaklmn+\kappa_{\mu})^2}\rhoklmn^{\,2}+4\frac{\kappa_{\mu}\alphaklmn\piklmn^{\,2}}{\alphaklmn+\kappa_{\mu}}-2\frac{\kappa_{\mu}\betaklmn\rhoklmn\cdot\piklmn}{\alphaklmn+\kappa_{\mu}}-5\frac{\kappa_{\mu}(\thetaklmn-\alphaklmn\lambdaklmn)}{\alphaklmn+\kappa_{\mu}}+2\Bigg\}\,G_{klmn}^{\,\mu},
\end{align}

\begin{align}
\bra{a_k\vec{b}_k\chi_k,a_l\vec{b}_l\chi_l}\op{r}^{\,2}G_{\mu}&(\op{r})\bar{S}_{12}(\vec{\op{p}}_{\Omega},\vec{\op{p}}_{\Omega})\ket{a_m\vec{b}_m\chi_m,a_n\vec{b}_n\chi_n}=\left(\frac{\kappa_{\mu}}{\alphaklmn+\kappa_{\mu}}\right)^2\cdot\nonumber \\
\Bigg\{S_{12}&(\rhoklmn\times\piklmn,\rhoklmn\times\piklmn)\bigg(\frac{\kappa_{\mu}^2\,\rhoklmn^{\,2}}{(\alphaklmn+\kappa_{\mu})^2}+5\frac{\alphaklmn\kappa_{\mu}}{\alphaklmn+\kappa_{\mu}}\bigg)\nonumber\\
+&S_{12}(\rhoklmn,\rhoklmn)\bigg(2\bigg(\frac{\kappa_{\mu}\,\rhoklmn\cdot\piklmn}{\alphaklmn+\kappa_{\mu}}\bigg)^2-\frac{3}{4}\frac{\kappa_{\mu}(\lambdaklmn\kappa_{\mu}+\thetaklmn)\,\rhoklmn^{\,2}}{(\alphaklmn+\kappa_{\mu})^2}\nonumber \\
&+4\frac{\kappa_{\mu}\betaklmn\,\rhoklmn\cdot\piklmn}{\alphaklmn+\kappa_{\mu}}+\frac{21}{4}\frac{\kappa_{\mu}(\thetaklmn-\alphaklmn\lambdaklmn)}{\alphaklmn+\kappa_{\mu}}+\frac{9}{4}\thetaklmn-\frac{9}{2}\bigg)\nonumber \\
+&S_{12}(\piklmn,\piklmn)\bigg(2\left(\frac{\kappa_{\mu}(\rhoklmn^{\,2})}{\alphaklmn+\kappa_{\mu}}\right)^2+13\frac{\alphaklmn\kappa_{\mu}\,\rhoklmn^{\,2}}{\alphaklmn+\kappa_{\mu}}+9\alphaklmn^2\bigg)\nonumber \\
-&S_{12}(\rhoklmn,\piklmn)\bigg(4\frac{\kappa_{\mu}^2\,\rhoklmn^{\,2}\rhoklmn\cdot\piklmn}{(\alphaklmn+\kappa_{\mu})^2}+\frac{5}{2}\frac{\kappa_{\mu}\betaklmn\,\rhoklmn^{\,2}}{\alphaklmn+\kappa_{\mu}}\nonumber \\
&+16\frac{\alphaklmn\kappa_{\mu}\,\rhoklmn\cdot\piklmn}{\alphaklmn+\kappa_{\mu}}+\frac{9}{2}\alphaklmn\betaklmn\bigg)\Bigg\}\,G_{klmn}^{\,\mu},
\end{align}
\begin{align}
\bra{a_k\vec{b}_k\chi_k,a_l\vec{b}_l\chi_l}\op{r}^{\,2}G_{\mu}(\op{r})\frac{1}{2}\Big[\vec{\op{L}}^{\,2}\bar{S}_{12}(\vec{\op{p}}_{\Omega},\vec{\op{p}}_{\Omega}) + \bar{S}_{12}(\vec{\op{p}}_{\Omega},\vec{\op{p}}_{\Omega}) \vec{\op{L}}^{\,2}\Big]\ket{a_m\vec{b}_m\chi_m,a_n\vec{b}_n\chi_n}=\left(\frac{\kappa_{\mu}}{\alphaklmn+\kappa_{\mu}}\right)^2\cdot \nonumber \\
\Bigg\{C^{S12LL}_{klmn}S_{12}(\rhoklmn\times\piklmn,\rhoklmn\times\piklmn)+C^{S12\rho\rho}_{klmn}S_{12}(\rhoklmn,\rhoklmn)\nonumber \\+C^{S12\pi\pi}_{klmn}S_{12}(\piklmn,\piklmn)+C^{S12\rho\pi}_{klmn}S_{12}(\rhoklmn,\piklmn)\Bigg\}\,G_{klmn}^{\,\mu},
\end{align}
\end{widetext}
with
\begin{widetext}
\begin{align}
C^{S12LL}_{klmn}=&\frac{\kappa_{\mu}^4\rhoklmn^{\,2}\left(\rhoklmn\times\piklmn\right)^2}{(\alphaklmn+\kappa_{\mu})^4}+\frac{3}{2}\frac{\kappa_{\mu}^3(\lambdaklmn\kappa_{\mu}+\thetaklmn)\,\left(\rhoklmn^{\,2}\right)^2}{(\alphaklmn+\kappa_{\mu})^4}-3\frac{\kappa_{\mu}^3\betaklmn\,\rhoklmn^{\,2}\rhoklmn\cdot\piklmn}{(\alphaklmn+\kappa_{\mu})^3}\nonumber \\
&+15\frac{\alphaklmn\kappa_{\mu}^3\,\rhoklmn^{\,2}\piklmn^{\,2}}{(\alphaklmn+\kappa_{\mu})^3}-9\frac{\alphaklmn\kappa_{\mu}^3\,\left(\rhoklmn\cdot\piklmn\right)^2}{(\alphaklmn+\kappa_{\mu})^3}\nonumber \\
&-\frac{\kappa_{\mu}^2\,\rhoklmn^{\,2}}{(\alphaklmn+\kappa_{\mu})^2}\bigg(24\frac{\kappa_{\mu}(\thetaklmn-\alphaklmn\lambdaklmn)}{\alphaklmn+\kappa_{\mu}}-\frac{21}{2}\thetaklmn-6\bigg)+42\frac{\alphaklmn^2\kappa_{\mu}^2\,\piklmn^{\,2}}{(\alphaklmn+\kappa_{\mu})^2}\nonumber\\
&-21\frac{\alphaklmn\kappa_{\mu}^2\betaklmn\,\rhoklmn\cdot\piklmn}{(\alphaklmn+\kappa_{\mu})^2}-\frac{\alphaklmn\kappa_{\mu}}{\alphaklmn+\kappa_{\mu}}\bigg(\frac{147}{2}\frac{\kappa_{\mu}(\thetaklmn-\alphaklmn\lambdaklmn)}{\alphaklmn+\kappa_{\mu}}-30\bigg),\nonumber
\end{align}

\begin{align}
C^{S12\rho\rho}_{klmn}=&2\frac{\kappa_{\mu}^4\,\left(\rhoklmn\cdot\piklmn\right)^2\left(\rhoklmn\times\piklmn\right)^2}{(\alphaklmn+\kappa_{\mu})^4}-\frac{3}{4}\frac{\kappa_{\mu}^3(\lambdaklmn\kappa_{\mu}+\thetaklmn)\,\left(\rhoklmn^{\,2}\right)^2\piklmn^{\,2}}{(\alphaklmn+\kappa_{\mu})^4}\nonumber \\
&+\frac{15}{4}\frac{\kappa_{\mu}^3(\lambdaklmn\kappa_{\mu}+\thetaklmn)\,\rhoklmn^{\,2}\left(\rhoklmn\cdot\piklmn\right)^2}{(\alphaklmn+\kappa_{\mu})^4}+6\frac{\kappa_{\mu}^3\betaklmn\,\rhoklmn\cdot\piklmn\left(\rhoklmn\times\piklmn\right)^2}{(\alphaklmn+\kappa_{\mu})^3}\nonumber \\
&-6\frac{\kappa_{\mu}^3\betaklmn\,\left(\rhoklmn\cdot\piklmn\right)^3}{(\alphaklmn+\kappa_{\mu})^3}+12\frac{\alphaklmn\kappa_{\mu}^3\,\left(\rhoklmn\cdot\piklmn\right)^2\piklmn^{\,2}}{(\alphaklmn+\kappa_{\mu})^3}\nonumber\\
&-\frac{3}{4}\frac{\kappa_{\mu}^2(\lambdaklmn\kappa_{\mu}+\thetaklmn)^2\,\left(\rhoklmn^{\,2}\right)^2}{(\alphaklmn+\kappa_{\mu})^4}+\frac{15}{2}\frac{\kappa_{\mu}^2\betaklmn(\lambdaklmn\kappa_{\mu}+\thetaklmn)\,\rhoklmn^{\,2}\rhoklmn\cdot\piklmn}{(\alphaklmn+\kappa_{\mu})^3}\nonumber\\
&+\frac{\kappa_{\mu}^2\,\rhoklmn^{\,2}\piklmn^{\,2}}{(\alphaklmn+\kappa_{\mu})^2}\bigg(\frac{51}{4}\frac{\kappa_{\mu}(\thetaklmn-\alphaklmn\lambdaklmn)}{\alphaklmn+\kappa_{\mu}}+\frac{21}{4}\thetaklmn-\frac{29}{2}\bigg)\nonumber\\
&-\frac{\kappa_{\mu}^2\left(\rhoklmn\cdot\piklmn\right)^2}{(\alphaklmn+\kappa_{\mu})^2}\bigg(\frac{165}{4}\frac{\kappa_{\mu}(\thetaklmn-\alphaklmn\lambdaklmn)}{\alphaklmn+\kappa_{\mu}}+\frac{225}{4}\thetaklmn-\frac{157}{2}\bigg)\nonumber\\
&+24\frac{\alphaklmn\kappa_{\mu}^2\betaklmn\,\rhoklmn\cdot\piklmn\piklmn^{\,2}}{(\alphaklmn+\kappa_{\mu})^2}\nonumber \\
&+\frac{\kappa_{\mu}(\lambdaklmn\kappa_{\mu}+\thetaklmn)\,\rhoklmn^{\,2}}{(\alphaklmn+\kappa_{\mu})^2}\bigg(\frac{27}{2}\frac{\kappa_{\mu}(\thetaklmn-\alphaklmn\lambdaklmn)}{\alphaklmn+\kappa_{\mu}}+\frac{15}{4}\thetaklmn-\frac{39}{4}\bigg)\nonumber\\
&-\frac{\kappa_{\mu}\betaklmn\,\rhoklmn\cdot\piklmn}{\alphaklmn+\kappa_{\mu}}\bigg(\frac{135}{2}\frac{\kappa_{\mu}(\thetaklmn-\alphaklmn\lambdaklmn)}{\alphaklmn+\kappa_{\mu}}+\frac{15}{2}\thetaklmn-37\bigg)\nonumber\\
&+\frac{\alphaklmn\kappa_{\mu}\,\piklmn^2}{\alphaklmn+\kappa_{\mu}}\bigg(\frac{69}{2}\frac{\kappa_{\mu}(\thetaklmn-\alphaklmn\lambdaklmn)}{\alphaklmn+\kappa_{\mu}}+15\thetaklmn-35\bigg)\nonumber \\
&-\frac{189}{4}\frac{\kappa_{\mu}^2(\thetaklmn-\alphaklmn\lambdaklmn)}{(\alphaklmn+\kappa_{\mu})^2}-\frac{21}{4}\frac{\kappa_{\mu}(\thetaklmn-\alphaklmn\lambdaklmn)}{\alphaklmn+\kappa_{\mu}}\bigg(5\thetaklmn-13\bigg)+\frac{27}{4}\thetaklmn-\frac{27}{2},\nonumber
\end{align}
\begin{align}
C^{S12\pi\pi}_{klmn}=&2\frac{\kappa_{\mu}^4\,\left(\rhoklmn\times\piklmn\right)^2\left(\rhoklmn^{\,2}\right)^2}{(\alphaklmn+\kappa_{\mu})^4}+3\frac{\kappa_{\mu}^3(\lambdaklmn\kappa_{\mu}+\thetaklmn)\left(\rhoklmn^{\,2}\right)^3}{(\alphaklmn+\kappa_{\mu})^4}-6\frac{\kappa_{\mu}^3\betaklmn\left(\rhoklmn^{\,2}\right)^2(\rhoklmn\cdot\piklmn)}{(\alphaklmn+\kappa_{\mu})^3}\nonumber \\
&+33\frac{\alphaklmn\kappa_{\mu}^3\,\left(\rhoklmn^{\,2}\right)^2\piklmn^{\,2}}{(\alphaklmn+\kappa_{\mu})^3}-21\frac{\alphaklmn\kappa_{\mu}^3\,\rhoklmn^{\,2}\left(\rhoklmn\cdot\piklmn\right)^2}{(\alphaklmn+\kappa_{\mu})^3}\nonumber \\
&-\frac{\kappa_{\mu}^2\,\left(\rhoklmn^{\,2}\right)^2}{(\alphaklmn+\kappa_{\mu})^2}\bigg(\frac{105}{2}\frac{\kappa_{\mu}(\thetaklmn-\alphaklmn\lambdaklmn)}{\alphaklmn+\kappa_{\mu}}-21\thetaklmn-16\bigg)\nonumber \\ 
&-42\frac{\alphaklmn\kappa_{\mu}^2\betaklmn\,\rhoklmn^{\,2}\rhoklmn\cdot\piklmn}{(\alphaklmn+\kappa_{\mu})^2}+117\frac{\alphaklmn^2\kappa_{\mu}^2\,\rhoklmn^{\,2}\piklmn^{\,2}}{(\alphaklmn+\kappa_{\mu})^2}-33\frac{\alphaklmn^2\kappa_{\mu}^2\left(\rhoklmn\cdot\piklmn\right)^2}{(\alphaklmn+\kappa_{\mu})^2}\nonumber\\
&-\frac{\alphaklmn\kappa_{\mu}\,\rhoklmn^{\,2}}{\alphaklmn+\kappa_{\mu}}\bigg(\frac{393}{2}\frac{\kappa_{\mu}(\thetaklmn-\alphaklmn\lambdaklmn)}{\alphaklmn+\kappa_{\mu}}-15\thetaklmn-74\bigg)\nonumber \\
&-30\frac{\alphaklmn^2\kappa_{\mu}\betaklmn\,\rhoklmn\cdot\piklmn}{\alphaklmn+\kappa_{\mu}}+60\frac{\alphaklmn^3\kappa_{\mu}\,\piklmn^{\,2}}{\alphaklmn+\kappa_{\mu}}-3\alphaklmn^2\bigg(35\frac{\kappa_{\mu}(\thetaklmn-\alphaklmn\lambdaklmn)}{\alphaklmn+\kappa_{\mu}}-9\bigg),\nonumber
\end{align}

\begin{align}
C^{S12\rho\pi}_{klmn}=&-4\frac{\kappa_{\mu}^4\,\rhoklmn^{\,2}\rhoklmn\cdot\piklmn\left(\rhoklmn\times\piklmn\right)^2}{(\alphaklmn+\kappa_{\mu})^4}-6\frac{\kappa_{\mu}^3(\lambdaklmn\kappa_{\mu}+\thetaklmn)\,\left(\rhoklmn^{\,2}\right)^2\rhoklmn\cdot\piklmn}{(\alphaklmn+\kappa_{\mu})^4}\nonumber \\
&-\frac{9}{2}\frac{\kappa_{\mu}^3\betaklmn\,\rhoklmn^{\,2}\left(\rhoklmn\times\piklmn\right)^2}{(\alphaklmn+\kappa_{\mu})^3}+12\frac{\kappa_{\mu}^3\betaklmn\,\rhoklmn^{\,2}\left(\rhoklmn\cdot\piklmn\right)^2}{(\alphaklmn+\kappa_{\mu})^3}\nonumber \\
&-48\frac{\alphaklmn\kappa_{\mu}^3\,\rhoklmn^{\,2}\,\rhoklmn\cdot\piklmn\,\piklmn^{\,2}}{(\alphaklmn+\kappa_{\mu})^3}+24\frac{\alphaklmn\kappa_{\mu}^3\left(\rhoklmn\cdot\piklmn\right)^3}{(\alphaklmn+\kappa_{\mu})^3}\nonumber\\
&-\frac{9}{2}\frac{\kappa_{\mu}^2\betaklmn(\lambdaklmn\kappa_{\mu}+\thetaklmn)\left(\rhoklmn^{\,2}\right)^2}{(\alphaklmn+\kappa_{\mu})^3}\nonumber \\
&+\frac{\kappa_{\mu}^2\,\rhoklmn^{\,2}\rhoklmn\cdot\piklmn}{(\alphaklmn+\kappa_{\mu})^2}\bigg(87\frac{\kappa_{\mu}(\thetaklmn-\alphaklmn\lambdaklmn)}{\alphaklmn+\kappa_{\mu}}+12\thetaklmn-68\bigg)\nonumber \\
&-\frac{69}{2}\frac{\alphaklmn\kappa_{\mu}^2\betaklmn\left(\rhoklmn\times\piklmn\right)^2}{(\alphaklmn+\kappa_{\mu})^2}+30\frac{\alphaklmn\kappa_{\mu}^2\betaklmn\left(\rhoklmn\cdot\piklmn\right)^2}{(\alphaklmn+\kappa_{\mu})^2}\nonumber \\
&-96\frac{\alphaklmn^2\kappa_{\mu}^2\,\rhoklmn\cdot\piklmn\piklmn^{\,2}}{(\alphaklmn+\kappa_{\mu})^2}+\frac{\kappa_{\mu}\betaklmn\,\rhoklmn^{\,2}}{\alphaklmn+\kappa_{\mu}}\bigg(48\frac{\kappa_{\mu}(\thetaklmn-\alphaklmn\lambdaklmn)}{\alphaklmn+\kappa_{\mu}}-\frac{15}{2}\thetaklmn-\frac{35}{2}\bigg)\nonumber \\
&+\frac{\alphaklmn\kappa_{\mu}\,\rhoklmn\cdot\piklmn}{\alphaklmn+\kappa_{\mu}}\bigg(201\frac{\kappa_{\mu}(\thetaklmn-\alphaklmn\lambdaklmn)}{\alphaklmn+\kappa_{\mu}}+60\thetaklmn-138\bigg)\nonumber\\
&-30\frac{\alphaklmn^2\kappa_{\mu}\betaklmn\,\piklmn^{\,2}}{\alphaklmn+\kappa_{\mu}}+\frac{3}{2}\alphaklmn\betaklmn\bigg(35\frac{\kappa_{\mu}(\thetaklmn-\alphaklmn\lambdaklmn)}{\alphaklmn+\kappa_{\mu}}-9\bigg).\nonumber
\end{align}
\end{widetext}

\end{document}